\documentclass{article} 
\usepackage[preprint]{colm2025_conference}

\usepackage{microtype}
\usepackage{hyperref}
\usepackage{url}
\usepackage{amsmath}
\usepackage{amssymb}
\usepackage{lineno}
\usepackage{enumitem}
\usepackage{booktabs}
\usepackage{cleveref} 
\usepackage{wrapfig,lipsum}
\usepackage{arydshln}
\usepackage{tcolorbox}
\usepackage{tabularx}

\setlist[itemize]{noitemsep}
\definecolor{darkblue}{rgb}{0, 0, 0.5}
\hypersetup{colorlinks=true, citecolor=darkblue, linkcolor=darkblue, urlcolor=darkblue}

\newcommand{\nparagraph}{\vspace{-0.9em}\paragraph}

\newcommand\codenew{\rule[-1pt]{0pt}{.7em}}
\newtcbox{\code}{on line, boxrule=0pt, boxsep=0pt, top=1.5pt,
left=2pt, bottom=1.5pt, right=2pt, colback=gray!25, colframe=white,
fontupper={\ttfamily\codenew}}

\newcommand{\benchname}{$L0$-Bench}

\title{$L$0-Reasoning Bench: Evaluating Procedural Correctness in \\Language Models via Simple Program Execution}

\author{
Simeng Sun, Cheng-Ping Hsieh, Faisal Ladhak, 
Erik Arakelyan, \\
\textbf{ Santiago Akle Serano}, \textbf{Boris Ginsburg}\\
NVIDIA \\
Santa Clara, CA 15213, USA \\
\texttt{\{simengs,bginsburg\}@nvidia.com}\\
}

\begin{document}

\ifcolmsubmission
\linenumbers
\fi

\maketitle

\begin{abstract}

Complex reasoning tasks often rely on the ability to consistently and accurately apply simple rules across incremental steps, a foundational capability which we term ``level-0'' reasoning.
To systematically evaluate this capability, we introduce \benchname, a language model benchmark for testing \emph{procedural} correctness -- the ability to generate correct reasoning processes, complementing existing benchmarks that primarily focus on \emph{outcome} correctness.
Given synthetic Python functions with simple operations, \benchname~grades models on their ability to generate step-by-step, error-free execution traces. 
The synthetic nature of \benchname~enables systematic and scalable generation of test programs along various axes (e.g., number of trace steps).
We evaluate a diverse array of recent closed-source and open-weight models on a baseline test set. All models exhibit degradation as the number of target trace steps increases, while larger models and reasoning-enhanced models better maintain correctness over multiple steps. 
Additionally, we use \benchname~to explore test-time scaling along three dimensions: input context length, number of solutions for majority voting, and inference steps. 
Our results suggest substantial room to improve ``level-0'' reasoning and potential directions to build more reliable reasoning systems. 

\end{abstract}
\section{Introduction}

Reasoning systems, by default, assume a fundamental prerequisite -- the consistent and accurate execution of simple rules. 
Whether performing multi-digit arithmetic or following a legal procedure, the ability to reliably execute predefined rules is essential, as even a small error can derail the process from an expected trajectory, regardless of how sophisticated the system might be in other respects.

We consider the ability to \emph{reliably execute simple rules across incremental steps} as the ``level-0'' reasoning ability, or a crucial form of sub-intelligence~\citep{morris2024levelsagioperationalizingprogress}.
This view also aligns with the recent discussion by~\citet{shalevshwartz2024artificialexpertintelligencepacreasoning} who formalize error-bounded precise reasoning in multi-step problem solving. However, modern neural language models are trained to approximate statistical patterns. \emph{Can these probabilistic models consistently and reliably execute discrete rules with guaranteed correctness?} 
To answer this question, we introduce \benchname, a synthetic benchmark to explore the feasibility of a language model functioning as a \emph{neural language computer}.
Specifically, given a simple program, a language model is graded on its ability to generate an execution trace step by step infallibly.

Focusing on programming languages offers several advantages. With precisely defined programs, \benchname~removes the inherent ambiguity of natural languages and isolates the reliable execution of simple rules from other reasoning processes (e.g., inferring general algorithms from observations). 
The callable functions also automatically produce execution traces, which allow for the evaluation of \emph{procedural} correctness (i.e., maintaining correctness across reasoning processes), an aspect often overshadowed by the simpler evaluation of \emph{outcome} correctness. 
Moreover, programs are easy to generate and can be systematically controlled.
In \benchname, we use generative grammar to construct simplified Python programs. 
By imposing multiple constraints on the grammar, we reduce the difficulty of executing each line of code, thereby avoiding unnecessary evaluation of operations specific to Python or arithmetic.
The rule-based generation also enables flexible control of various axes (e.g., number of trace steps, program length) while being significantly more cost-effective than LM-based data curation. 
Furthermore, the synthetic nature mitigates concerns about data leakage/memorization and allows for scalable generation of test instances.

\benchname~relates to prior studies on algorithmic tasks~\citep{lee2024teaching,markeeva2024clrstextalgorithmicreasoninglanguage}, such as large-number multiplication~\citep{deng2024explicit}. 
However, \benchname~differs in providing a general framework where the input program is not limited to a single predefined routine; instead, \benchname~is able to admit any task that can be formulated as a program. 
While external tools naturally excel at executing discrete rules, we argue that the ability to perform simple operations step-by-step like a Turing Machine, without external tools, is still critical. It is essential not only for reasoning tasks that rely on basic deduction, but also for facilitating robustness, interpretability, and faithful alignment with human intents.

We evaluate 20 models with \benchname, including closed-source models, open-weight general-instruct models (ranging from 7B to 405B), and reasoning-enhanced models that generate ``think'' patterns, such as DeepSeek-R1. Experimental results demonstrate the advantages of larger model sizes and reasoning enhancements in achieving better procedural correctness. However, all models exhibit degradation as the target number of trace steps increases. We further leverage \benchname~as a constrained environment to explore test-time scaling along three dimensions: (1) scaling input context by increasing step-by-step demonstrations, (2) scaling width by increasing the number of solutions for majority voting, and (3) scaling inference steps by enabling long chain-of-thought. 
While each dimension independently improves performance, our results reveal key limitations: 
scaling many-shot demonstrations yields diminishing returns, and sometimes degrades performance even for \emph{long-context} models; majority voting plateaus as the number of voter grows; and long chain-of-thought reasoning increases latency without consistently benefiting all model sizes.
Nonetheless, our scaling analysis demonstrates the benefits of scaling alone all three scaling dimensions and suggests ample room for improvement in current models. 

Our contributions are as follows:
\begin{itemize}[leftmargin=*]
    \item We introduce \benchname, a synthetic benchmark for evaluating procedural correctness in neural language models via program execution. \benchname~isolates step-by-step rule-execution from other reasoning processes (e.g., induction and search) and provides a controlled, verifiable evaluation framework.
    \item By evaluating 20 recent language models, we reveal their limitations in maintaining correctness across simple multi-step procedures, and show the benefits of larger model sizes and reasoning enhancements with long chain-of-thought.
    \item We further demonstrate the benefits of test-time scaling along three dimensions (input demonstrations, majority voting, and inference steps) and discuss the trade-off between performance and efficiency.
\end{itemize}
Overall, our findings reveal key limitations of current LLMs in long step-by-step rule-
execution and suggest potential directions toward more reliable reasoning systems.\footnote{We release code at \url{https://github.com/SimengSun/L0-reasoning-bench}.}

\section{L-0 Benchmark} \label{sec:benchmark}

\subsection{Benchmark construction}

Program execution can be viewed as a special form of deduction. During execution time, simple predefined rules are applied step-by-step, moving the execution process from one state to the next deterministically. For instance, \code{list.pop()} requires applying the \emph{general rule} of ``popping an item'' to a list \emph{instance}, such as \code{list = [1,2,3,4]}.
Evaluating a model's ability to execute simple programs therefore can serve as a test of its deductive capability -- a necessary but not sufficient condition for complex reasoning processes. We introduce \benchname~to test the deductive capability in neural language models via program execution, i.e., the ability to follow unambiguously defined routines step-by-step without deviation, akin to a neural language computer.

In \benchname, we construct synthetic Python programs using generative grammar.
To avoid testing Python-specific operations or evaluations of complex expressions, we impose multiple constraints on the generative grammar, while preserving fundamental elements of conventional programming languages: (a) simple arithmetic and comparison operations, (b) control (conditional branch and loops), and (c) memory access and operations. The grammar hence covers diverse procedures that require modifying execution states (e.g., append an item to a list) and strictly adhering to the control flow (e.g., exit loops on termination), which are absent in simpler rule-based tasks, such as variable tracking~\citep{hsieh2024ruler}.
\begin{figure}[!t]
    \centering
\includegraphics[width=\linewidth]{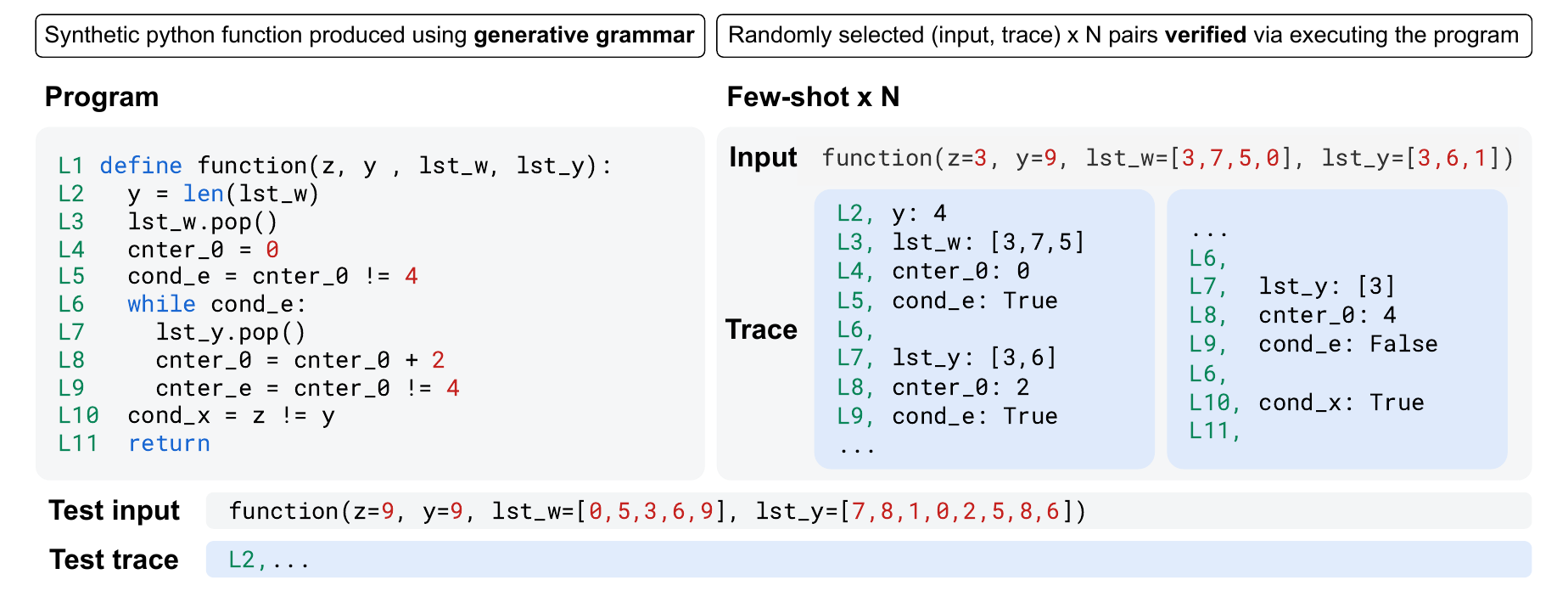}
    \caption{An illustrative example in \benchname. The prompt includes a program, multiple few-shot execution demonstrations, and a test input. Programs are generated using a constrained generative grammar that defines a simplified Python subset (see  Appendix~\ref{sec:appendix_grammar}). Inputs are randomly sampled and verified by executing the Python function to ensure error-free evaluation.}
    \label{fig:figure_1}
\end{figure}
\nparagraph{Program generation.} To reduce the difficulty of executing each line of code, we impose multiple constraints on the production rules used to generate the test programs. A detailed grammar description is provided in Appendix~\ref{sec:appendix_grammar}, with  main constraints summarized below:
\begin{itemize}[leftmargin=*]
\small \vspace{-1em}
    \item The generated program is a standalone Python function.
    \item The variables are one of the following: integer, list of integers, and boolean value. 
    \item We restrict basic binary operations to \code{+}, \code{-}, \code{==}, and \code{!=}, as models struggle to reliably execute other operations such as \code{*}, \code{\%}, and \code{<}.
    \item Complex single-line expressions are disabled to ensure the simplicity of transition between lines. For instance, we do not allow evaluation of long expressions such as \code{(x + lst[5] - y - 6)}.
    \item Context-sensitive rules are added to enforce terminable while loops. 
\end{itemize}
\nparagraph{Verified inputs and traces.} We randomly generate values for the function input arguments and skip inputs that encounter execution errors (e.g., pop an item from an empty list). For each program, we generate and verify multiple sets of (input, trace) pairs, which can serve as demonstrations for the expected trace format. Each trace step consists of a line number, a variable name, and its updated value, if applicable. If no variable value is updated, e.g., the line of \code{if condition:} or \code{while condition:}, only the line number is required. 

\nparagraph{Example format.} We provide $N$ few-shot demonstrations specific to the test program for guiding the generation. Formally, let $\mathcal{F}$ denote the string of the provided program, $x_i$ be the string of a demonstration input, and $y_i$ be its corresponding output trace consisting of multiple trace lines. Let $x^*$ and $y^*$ represent the test input and expected output trace, respectively. The task input is structured as $\{\mathcal{F}, x_1, y_1, x_2, y_2, \dots, x^*\}$, with the expected output being $\{y^*\}$. The prompt template and example are provided in Appendix~\ref{sec:prompt_example}. An illustrative example is shown in Figure~\ref{fig:figure_1}.

\subsection{Evaluation Metrics}
We evaluate whether a model can generate the correct execution trace, both from beginning to end (whole-trace accuracy) and partially (steps to the first error):
\begin{itemize}[leftmargin=*]
\vspace{-0.5em}
\item \textbf{Whole-trace accuracy}: Since all traces begin with a fixed format, we align the model response with the start of the ground-truth trace, and then evaluate whether the entire generated sequence exactly matches the expected ground-truth trace.  
\item \textbf{Steps to the first error}: We also evaluate the partial correctness by counting the number of correct steps before the first error appears. We do not evaluate correctness beyond the first error, as the model may enter an entirely incorrect branch or prematurely exit loops, making subsequent steps incorrect. 
\end{itemize}
\vspace{-0.5em}
The two metrics are not strictly correlated: a model with a low \textbf{whole-trace accuracy} can still have a high \textbf{steps to the first error} if errors occur primarily at the end of the trace.

\section{Experimental Setup}

\nparagraph{Data.} \benchname~is designed to flexibly generate data based on specified configurations. We construct a base set of \benchname~by enforcing multiple constraints described in Appendix~\ref{sec:appendix_grammar}. These constraints can be adjusted to increase difficulty, such as allowing for longer expressions at each line, increasing scope depth, etc. We group generated examples into four bins of \emph{short}, \emph{medium}, \emph{long}, and \emph{extra long} traces. Each bin contains 500 examples with the average number of steps  $\{13, 80, 164, 246\}$ respectively.

\nparagraph{Prompt Format \& Majority Voting Setup.} We include four (input, trace) demonstrations specific to the test program in the context. Both evaluation metrics are reported for the majority-voted responses (\emph{majvote@31}) across 31 parallel samplers as well as the accuracy of having at least one whole correct trace (\emph{pass@31}). The majority-voted trace is the most frequent whole trace after aligning the voters' responses based on the fixed ground-truth beginning pattern (``\emph{L2,}''). The ``\emph{Single Attempt}'' results are the average of 31 independent runs, where each voter randomly draws four distinct demonstrations from a pool of 64 exempla pre-generated for each test program.

\nparagraph{Models.} We evaluate 20 models, covering both closed-source and open-weight models ranging from 7B to 405B, including general instruct models as well as reasoning-enhanced models such as DeepSeek-R1 and R1-Distilled Qwen models. For all models, we use greedy decoding, except for R1 and R1-distilled models, which are sampled with the recommended temperature of 0.6. For models that generate long Chain-of-Thought outputs, we cap the maximum number of generated tokens at 20 times the length of the ground-truth trace. To disable the long reasoning pattern among these models, we explicitly append an answer prefix to the chat template to bypass the generation of \texttt{<think>} token.

\section{Main Results}

\nparagraph{\benchname~\& general-instruct models.} One potential concern of \benchname~is that the focus on programming language may pose challenges for general-instruct models that lack code-specific training. 
Results in Table~\ref{tab:main_res_tab} reveal that the general-instruct model \texttt{Qwen2.5-32B-Instruct} in fact performs better than its same-size \texttt{Qwen2.5-Coder} model, suggesting that \benchname~can be a viable test for general-instruct models.

\nparagraph{Benefits of larger model sizes and ``think'' patterns.} The column \emph{thinking mode on} is set to ``Y'' in Table~\ref{tab:main_res_tab} for runs that produce long chain-of-thought ``think'' patterns. 
Among ``non-thinking'' systems ranging in \{7B, 14B, 32B, 72B, 405B\}, we observe the general trend: larger models achieve higher accuracy in both single-attempt evaluations and majority-voted whole-trace accuracy, indicating the benefits of scaling base model parameters. While model size generally correlates with better performance, models that generate intermediate chain-of-thought steps can outperform much larger ``non-thinking'' counterparts. For instance, \texttt{DeepSeek-R1-Distilled-Qwen-14B} outperforms the original 72B Qwen2.5 as well as the 29 times larger Llama. However, the improved performance comes at the cost of longer inference time. We discuss the performance-efficiency tradeoff in a later section.

\begin{table}[!t]
\centering
\scalebox{0.87}{
\begin{tabular}{@{}lc|cc|ccc@{}}
\toprule
& & \multicolumn{2}{c|}{\textbf{\begin{tabular}{@{}c@{}}Steps to Err. $\uparrow$  \\ (target \# steps = 125.8)\end{tabular}}} & \multicolumn{3}{c}{\textbf{Trace Acc. (\%)}} \\ 
\textbf{Model Name} & \begin{tabular}{@{}c@{}}Thinking\\ mode on\end{tabular} & {\begin{tabular}{@{}c@{}}Single\\ Attempt\end{tabular}}  & {\begin{tabular}{@{}c@{}}majvote \\ @ 31\end{tabular}} & {\begin{tabular}{@{}c@{}}Single\\ Attempt\end{tabular}} & {\begin{tabular}{@{}c@{}}majvote \\ @ 31\end{tabular}} & {\begin{tabular}{@{}c@{}}pass \\ @ 31\end{tabular}} \\ 
\midrule
\textbf{Closed-source models}\\
\hspace{1em} o1 & Y & 122.2 & - & 92.0 & - & - \\
\hspace{1em} o3-mini & Y & 92.8 & - & 60.6 & - & - \\
\hspace{1em} gpt-4o-2024-11-20 & N& 112.5 & -& 76.4   & -& -\\
\midrule
\hspace{1em} claude-3-7-sonnet & Y& 118.6 & -& 95.1   & -& -\\ 
\hspace{1em} claude-3-7-sonnet & N& 114.0 & -& 86.1   & -& -\\
\hspace{1em} claude-3-5-sonnet-20241022& N& 108.5 & -& 83.3   & -& -\\
\midrule \midrule
\textbf{Open-weight models}\\
\hspace{1em} QwQ-32B   & Y& 120.0 & 124.9& 86.6  & 96.1 & 99.8 \\ 
\hspace{1em} Deepseek-R1   & Y& 115.7 & 121.4& 91.5  & 97.4 & 97.9 \\ 
\midrule
\hspace{1em} DeepSeek-R1-Distill-Llama-70B & Y& 104.4 & 116.3& 62.0  & 79.6 & 96.6 \\
\hspace{1em} Meta-Llama-3.1-405B-Instruct  & N& 99.1  & 106.2& 63.0  & 73.1 & 93.2 \\
\hspace{1em} Meta-Llama-3.1-70B-Instruct   & N& 82.8  & 93.9 & 42.4  & 54.7 & 85.9 \\ 
\hspace{1em} Meta-Llama-3.1-8B-Instruct & N& 33.2  & 40.7 & 6.5   & 11.5 & 33.9 \\
\midrule 
\hspace{1em} DeepSeek-R1-Distill-Qwen-32B  & Y& 114.4 & 123.4& 80.4  & 94.9 & 99.8 \\
\hspace{1em} DeepSeek-R1-Distill-Qwen-14B  & Y& 99.3  & 118.3& 56.8  & 83.5 & 97.8 \\ 
\hspace{1em} DeepSeek-R1-Distill-Qwen-7B\footnotemark[2]   & Y& 8.7   & 11.7 & 3.8   & 11.2 & 28.8 \\ 
\hspace{1em} Qwen2.5-72B-Instruct  & N& 94.3  & 102.6& 55.2  & 66.9 & 89.9 \\
\hspace{1em} Qwen2.5-32B-Instruct  & N& 88.3  & 101.8& 50.4  & 66.0 & 90.0 \\
\hspace{1em} Qwen2.5-Coder-32B-Instruct& N& 81.4  & 92.4 & 44.0  & 56.8 & 84.4 \\
\hspace{1em} Qwen2.5-14B-Instruct  & N& 60.9  & 73.4 & 23.7  & 35.7 & 73.7 \\
\hspace{1em} Qwen2.5-Coder-7B-Instruct & N& 52.3  & 68.4 & 17.1  & 32.4 & 62.5 \\
\hspace{1em} Qwen2.5-7B-Instruct   & N& 41.2  & 54.2 & 11.3  & 22.8 & 44.1 \\
\bottomrule
\end{tabular}}
\caption[]{
We evaluate 20 recent models on \benchname. Results are averaged over four data splits (short, medium, long, extra-long traces), with \emph{Single Attempt} results based on the average of 31 independent runs. Larger models, reasoning with longer chain-of-thought, and majority voting over parallel runs improve performance on \benchname. The large gap between \emph{pass@k} and \emph{single attempts} across all models indicates significant room for improvement. Details about tested models are provided in Table~\ref{tab:model_link}. }
\label{tab:main_res_tab}
\end{table}
\footnotetext[2]{Model often fails to exit the ``thought'' process with naive budget forcing~\citep{muennighoff2025s1} (maximum generation length is set to 20 times the target length), thus performing worse overall.}

\begin{figure}[!t]
    \centering
    \includegraphics[width=0.92\linewidth]{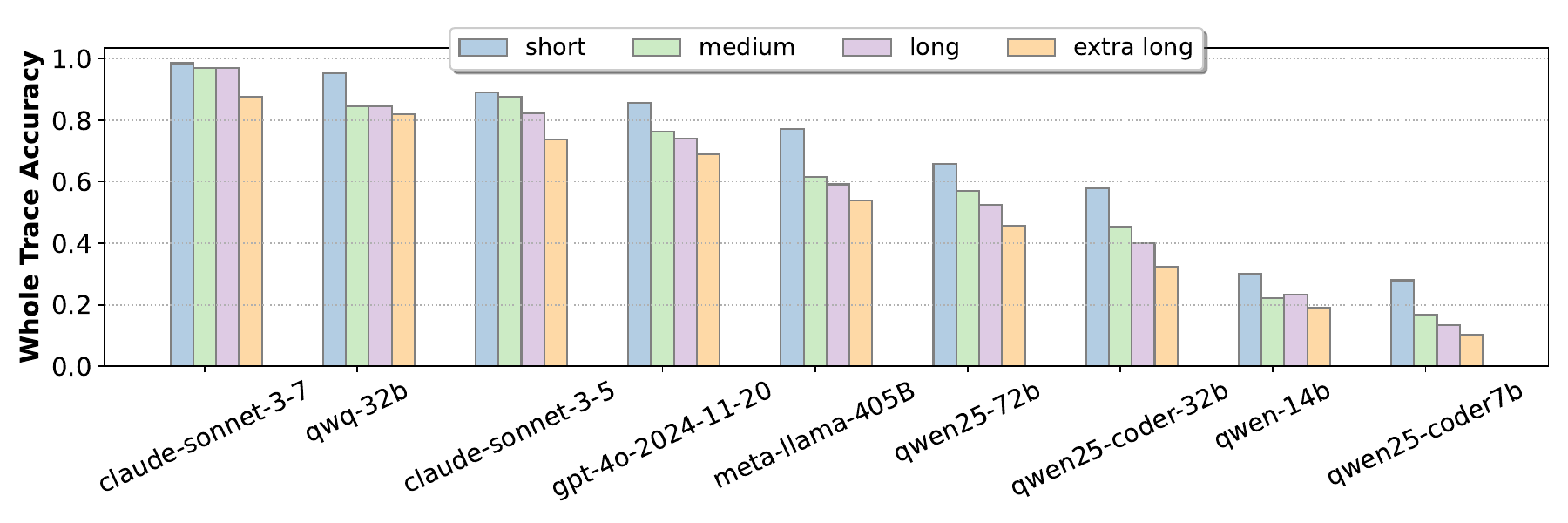}
    \caption{Model performance generally declines as the number of trace steps increases. Open-weight models in the figure are instruct version.}
    \label{fig:perf_by_split}
\end{figure}

\nparagraph{Models fail at longer traces more often.} While programs in \benchname~are simplified to allow for only basic operations, 
none of the models, including reasoning-enhanced models, is able to consistently execute the basic operations as the number of ground-truth trace steps increases (Figure~\ref{fig:perf_by_split}). For instance, while \texttt{QwQ-32B} achieves over 95\% whole-trace accuracy on the shortest split, its performance drops to 82\% on the longest split. This limitation persists despite the constrained set of simple operations and the total number of tokens in the prompt being far below the claimed effective context size.

\nparagraph{Majority voting boosts performance, yet falls far behind Pass@k.} Consistent with other works, majority voting over multiple solutions brings significant gains for all data splits. The gains are even more pronounced for \emph{pass@31}, a soft upper-bound that models are yet to reach. 
Notably, reasoning-enhanced models, such as \texttt{QwQ-32B}, demonstrate much smaller gaps between majority-voted and \emph{pass@k} accuracy, however, the gaps are still non-marginal, suggesting promising directions to improve the latest models. 

\nparagraph{Steps To Error provides a more fine-grained evaluation.} Generally, models achieve high whole-trace accuracy when their average \emph{Steps to Error} is also high. However, we observe discrepancies between the whole-trace and partial accuracy for certain models. For instance, despite achieving an average $\sim$112.5 steps to the first error, \texttt{gpt-4o} scores below $80$ in whole-trace accuracy. This aligns with manual examination which reveals that the model often fails towards the end of the trace by skipping steps.

\section{Scaling Test-Time Compute}

We employ \benchname~as a controllable environment to systematically explore three dimensions of test-time scaling: (1) \textbf{scaling input context} by increasing the number of in-context demonstrations, (2) \textbf{scaling width} by increasing the number of solutions for majority voting, and (3) \textbf{scaling inference steps} by enabling long chain-of-thought.

\nparagraph{Scaling input context.} Learning from in-context demonstrations can be viewed as ``optimization'' through forward passes~\citep{icl_grad}, or inference-time training. 
We scale the number of in-context demonstrations up to 64 for the 70B Llama model. Table~\ref{tab:scaling_input} shows that performance consistently improves as the number of demonstrations increases from 2 to 32 except for the extra-long split. Scaling beyond 32-shot brings little gains or even hurts performance, despite the input length being much shorter than the supported context of 128K tokens. This is not ideal as all demonstrations provide step-by-step execution of the same test program, thus immediately relevant to producing the target trace. This many-shot setting of \benchname~can therefore serve as an alternative to existing long-context benchmarks that often emphasize on the capability to retrieve relevant signals from irrelevant noises.
\begin{table}[]
\centering
\scalebox{0.85}{
\begin{tabular}{@{}lccccccc@{}}
\toprule
 & \textbf{\# Target Tokens} & \textbf{2-shot}   & \textbf{4-shot}   & \textbf{8-shot}   & \textbf{16-shot}  & \textbf{32-shot}  & \textbf{64-shot}  \\
 \midrule
\textbf{\# Input tokens}& \multicolumn{1}{l}{}   & \multicolumn{1}{l}{} & \multicolumn{1}{l}{} & \multicolumn{1}{l}{} & \multicolumn{1}{l}{} & \multicolumn{1}{l}{} & \multicolumn{1}{l}{} \\
\hspace{1em} Short trace & 0.1K   & 1K   & 1.5K & 2.7K & 5K   & 9.5K & 19K  \\
\hspace{1em} Medium trace& 0.8K   & 2.3K & 4.1K & 7.8K & 15K  & 30K  & 59K  \\
\hspace{1em} Long trace  & 1.9K   & 4.4K & 8.4K & 16K  & 32K  & 64K  & {\color[HTML]{999999}129K} \\
\hspace{1em} Extra long trace & 3.1K   & 7.0K & 13K  & 27K  & 53K  & 100K & {\color[HTML]{999999}208K} \\ 
\midrule
\textbf{Pass@1 Trace Acc. (\%)} & \multicolumn{1}{l}{}   & \multicolumn{1}{l}{} & \multicolumn{1}{l}{} & \multicolumn{1}{l}{} & \multicolumn{1}{l}{} & \multicolumn{1}{l}{} & \multicolumn{1}{l}{} \\
\hspace{1em} Short trace & 0.1K   & 49.1$_{\pm 2.0}$ & 58.2$_{\pm 1.5}$ & 66.7$_{\pm 1.4}$ & 71.1$_{\pm 1.4}$ & 75.5$_{\pm 1.3}$ & 75.3$_{\pm 1.1}$\\
\hspace{1em} Medium trace & 0.8K   & 32.4$_{\pm 1.2}$ & 40.4$_{\pm 1.6}$ & 48.0$_{\pm 1.5}$& 53.3$_{\pm 1.5}$& 55.3$_{\pm 1.1}$& 49.2$_{\pm 1.4}$\\
\hspace{1em} Long trace  & 1.9K   & 28.1$_{\pm 1.2}$ & 36.3$_{\pm 1.5}$ & 43.1$_{\pm 1.8}$ & 46.6$_{\pm 1.4}$& 39.9$_{\pm 1.1}$& {\color[HTML]{999999} 23.5$_{\pm 0.9}$} \\
\hspace{1em} Extra long trace & 3.1K   & 28.0$_{\pm 1.4}$ & 34.6$_{\pm 1.6}$& 38.6$_{\pm 1.4}$& 36.9$_{\pm 1.7}$& 23.4$_{\pm 1.1}$ & {\color[HTML]{999999} 1.5$_{\pm 0.4}$} \\
\bottomrule
\end{tabular}
}
\caption{We scale up to 64 few-shot demonstrations specific to the input test programs for \texttt{Meta-Llama-3.1-70B-Instruct}, adopting a many-shot setting where all demonstrations are directly relevant, unlike many long-context benchmarks that emphasize extracting signals from distractors. 
While including more step-by-step demonstrations in the context improves performance for short traces, performance degrades for longer traces even before reaching the 128K context limit. The results evaluated on context longer than 128K tokens are in \color[HTML]{999999}{gray}.}
\label{tab:scaling_input}
\end{table}

\begin{figure}
    \centering
    \includegraphics[width=0.5\linewidth]{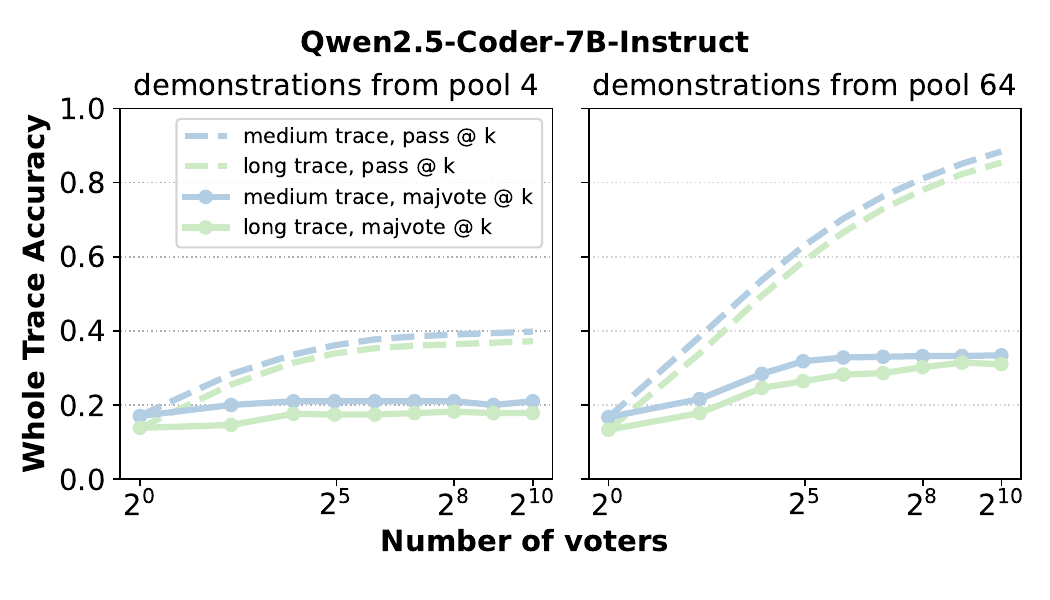}\includegraphics[width=0.5\linewidth]{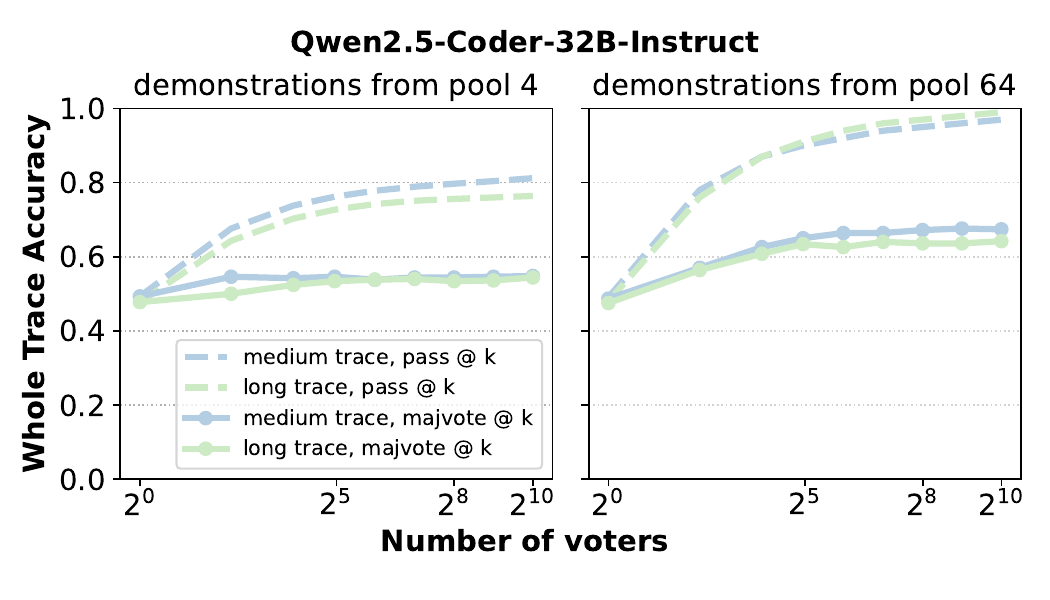}
    \caption{We increase the number of voters up to 1024 to study majority voting at scale. 
    Each voter either permutes the same set of four demonstrations or randomly selects four distinct few-shot demonstrations from a pool of 64. Access to a larger, more diverse demonstration pool improves majority-voted performance and pass@K. The significant gap between pass@K and majority-voted performance indicates large room for models to reach their full potential.
    }
\label{fig:scaling_width}
\end{figure}

\nparagraph{Scaling width.}  To explore the potential of majority voting at scale, we increase the number of solutions to 1K for two models from the Qwen series under two settings in Figure~\ref{fig:scaling_width}: (1) each voter permutes the same fixed set of four-shot demonstrations of the test program, and (2) each voter independently selects four distinct demonstrations from a pool of 64 w.r.t the test program, same as the main results.
In both cases, whole-trace accuracy improves with the number of solutions, with a diminishing return beyond 32$\sim$64. Notably, drawing demonstrations from a diverse pool yields larger gains, with the small 7B model showing promising \emph{pass@k} with 1K solutions. These results imply the presence of correct traces in models' hypothesis space, yet models struggle to reliably and effectively leverage in-context demonstrations with low sample complexity during inference-time training.

\nparagraph{Scaling inference steps.} Recent works suggest that scaling scratchpad length helps overcome architectural limitations and empowers Transformer to solve tasks previously unsolvable with limited depth~\citep{li2024chain}. Our results confirm the benefit of generating a long chain-of-thought sequence prior to providing the final answer on \benchname. Specifically, results in Table 3 reveal large performance drop when the ``think'' patterns in reasoning-enhanced models are bypassed.\footnote[3]{We modify the chat template \texttt{<|Assistant|>\ The execution trace is:}\texttt{\textbackslash n} to bypass the \texttt{<think>}.} For instance, disabling ``thinking'' in \texttt{DeepSeek-R1-Distill-Qwen-32B} causes a drop in whole-trace accuracy of $\sim$35 points, while also harming the soft upper-bound \emph{pass@k} performance. This indicates the benefits of step-by-step reasoning and self-correction patterns for maintaining correctness over multiple steps, as shown in an example \texttt{QwQ-32B} response in Appendix~\ref{tab:qwq_example_0}.

\nparagraph{Scaling along all dimensions \& limitations.} Combining the above three test-time scaling dimensions often leads to additional gains. For instance, the reasoning-enhanced 70B Llama significantly outperforms the original one without ``think'' patterns. After providing more demonstrations and aggregating over multiple voters (Table 4), we observe significant gains in whole-trace accuracy (\textbf{65}$\rightarrow$\textbf{88}). However, scaling along each dimension incurs different costs. The long-CoT model can be slow at inference,\footnote[4]{For example, \texttt{DeepSeek-R1-Distill-Qwen-32B} takes $29$ min to complete the \emph{short} split, whereas \texttt{Qwen2.5-32B-Inst} takes around $3$ min, both running on 4 A100 GPUs with the same serving setup.} whereas taking majority vote over solutions generated \emph{in parallel} avoids high latency, only when sufficient compute resources are available. Additionally, scaling the number of demonstrations can be promising for long-context models, however, access to a large pool of demonstrations may not always be available. We leave a systematic investigation of these trade-offs for future work.

\begin{table}[!t]
\begin{minipage}{0.62\textwidth}
\centering

\resizebox{0.99\linewidth}{!}{
\begin{tabular}{@{}lc|ccc@{}}
\toprule
& \multicolumn{1}{l}{}& \multicolumn{3}{c}{\textbf{Trace Acc. (\%)}}\\ 
\textbf{Model Name} & \begin{tabular}[c]{@{}c@{}}Thinking\\ mode on\end{tabular} & \begin{tabular}[c]{@{}c@{}}Single\\ Attempt\end{tabular} & \begin{tabular}[c]{@{}c@{}}maj vote\\ @ 31\end{tabular} & \begin{tabular}[c]{@{}c@{}}pass \\ @ 31\end{tabular} \\ \midrule
QwQ-32B & Y &  86.6	& 96.1 & 99.8\\
QwQ-32B & N & 23.3	& 64.0 & 85.4 \\
R1-Distill-Qwen-32B & Y & 80.4	& 94.9 & 99.8 \\
R1-Distill-Qwen-32B & N & 45.3 & 60.1 & 86.6 \\
Qwen2.5-32B-Inst & N & 50.4	& 66.0 &	90.0 \\
\bottomrule
    \end{tabular}}
\label{tab:scale_inference}
\caption{\small Disabling long chain-of-thought in reasoning-enhanced models leads to large performance drop.}
\end{minipage}
\hfill
\begin{minipage}{0.35\textwidth}
\centering
\resizebox{0.99\linewidth}{!}{
\begin{tabular}{@{}c|ccc@{}} 
\toprule
\multicolumn{1}{l}{}& \multicolumn{3}{c}{\textbf{Trace Acc. (\%)}}\\ 
\begin{tabular}[c]{@{}c@{}}\# few \\  shot\end{tabular} & \begin{tabular}[c]{@{}c@{}}Single\\ Attempt\end{tabular} & \begin{tabular}[c]{@{}c@{}}maj vote\\ @ 31\end{tabular} & \begin{tabular}[c]{@{}c@{}}pass \\ @ 31\end{tabular} \\ \midrule
2 & \textcolor{red}{\textbf{65.5}} & 83.8 & 98.2\\
4 & 72.9 & 87.6 & 98.6\\
8 & 75.1 & \textcolor{blue}{\textbf{88.2}} & 99.2\\
16& 74.1 & 86.6 & 98.6\\
32& 69.5 & 81.6 & 97.4 \\
 \bottomrule
\end{tabular}}
\label{tab:scale_all}

\caption{\small R1-Distill-Llama-70B on the \emph{medium} split. }
\end{minipage}
\end{table}
\section{Ablation \& Analysis}
\begin{figure}[!t]
    \centering
    \includegraphics[width=0.42\linewidth]{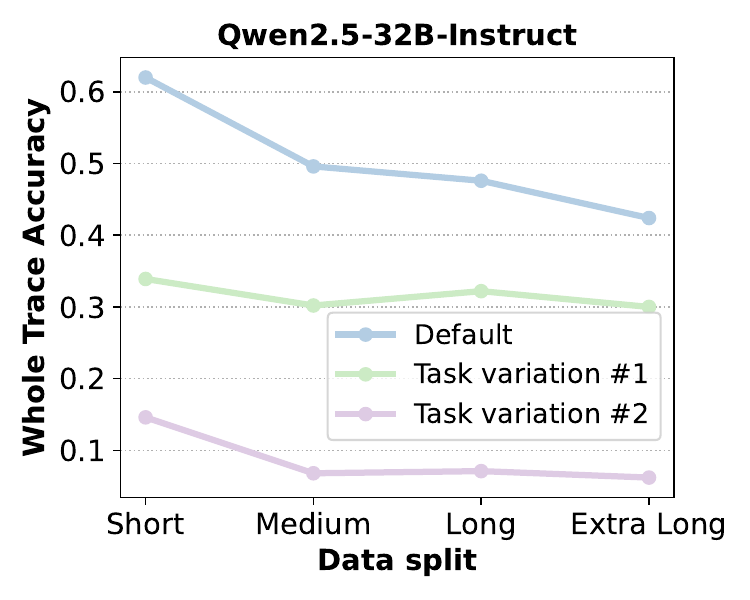}\includegraphics[width=0.42\linewidth]{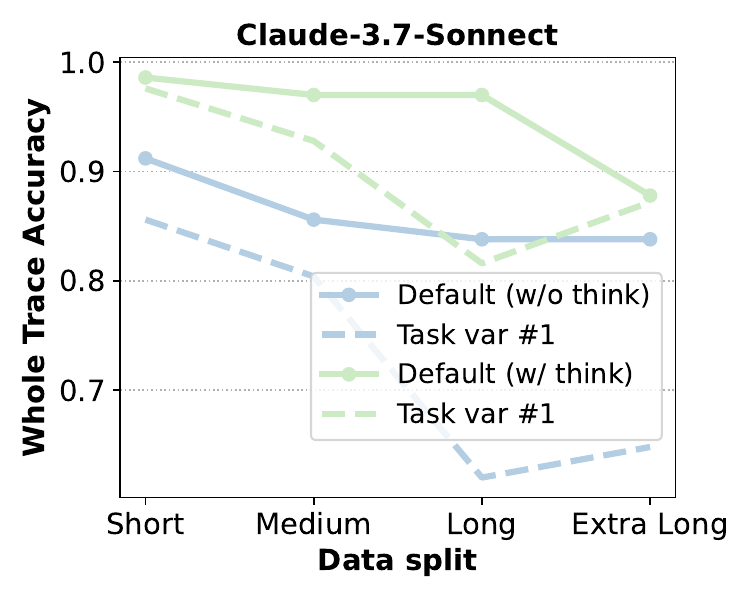}
    \caption{\textbf{(Left)} The difficulty of \emph{default} tasks in \benchname~can be increased by : (\emph{Task variation \#1}) using irrelevant (program, input, trace) as few-shot examples, or (\emph{Task variation \#2}) removing the program, forcing the model to infer the algorithm purely from step-by-step demonstrations. We use the simplest setup to test the 0-level reasoning. \textbf{(Right)} Even the strongest closed-source model Claude-3.7-Sonnet shows performance degradation with the modification of task setup, especially when extended thinking is disabled. }
    \label{fig:task_vars}
\end{figure}

\nparagraph{Task variation \# 1: Alternative Programs as Demonstrations} The default setup in \benchname~uses (input, trace) pairs of the same test program as demonstrations. This setup is much easier than a variation setup where demonstrations are other programs, and models no longer have access to the step-by-step demonstrations on how to execute the test algorithm. That is, the prompts are structured as $\{\mathcal{F}_1, x_1, y_1, \mathcal{F}_2, x_2, y_2, \dots, \mathcal{F}_*, x^*\}$, following the notation used in \S~\ref{sec:benchmark}. 
Figure~\ref{fig:task_vars} shows performance degradation for both \texttt{Qwen2.5-32B-instruct} and the performant closed-source model \texttt{claude-sonnet-3.7}. We chose the easier setup on purpose to avoid distractors in the prompt in \benchname.

\nparagraph{Task variation \# 2: Removing Input Program -- a  Transduction Task.} We test whether in-context step-by-step demonstrations alone suffice for generating execution traces by removing the input program from the prompt, a transduction setup similar to ARC-AGI~\citep{chollet2019measureintelligence,chollet2024arc}. Figure~\ref{fig:task_vars} (left) indicates that the 32B Qwen model achieves non-zero accuracy even without the input program, implying occasionally correct pattern inference from the provided (input, trace) pairs. We focus on the simpler setting in this work, leaving the much harder transduction task for future exploration.

\nparagraph{Larger models tend to spend less time ``thinking.''} We count the number tokens enclosed by \texttt{<think>} and \texttt{</think>} in five reasoning-enhanced models. The larger 70B generates on average 1365 thought tokens, significantly fewer tokens than the smaller Qwen models, which generate on average \{3344, 2241, 1722\} for sizes of \{7B, 14B, 32B\}, respectively. 
 Despite the relationship observed in R1-distilled models, \texttt{QwQ-32B} is more verbose than the other models, generating over 6K thought tokens on average.

\nparagraph{Common Failure Modes} Manual inspection of errors reveals three common failure modes. (1) \textbf{Incorrect item counting}: Even closed-source models struggle to reliably copy a list from a previous step to the next during operations like \texttt{pop} or \texttt{append}. Often, items are incorrectly repeated in the list, or completely deleted even when the list is not long. (2) \textbf{Mis-evaluation of simple expression}: Despite the simplified binary operations in \benchname, small models still fail to reliably evaluate simple expressions, such as \code{10 == 10}, thus leading to incorrect branching decisions for \emph{if} blocks sometimes. (3) \textbf{Skipped steps}: models sometimes incorrectly skip trace steps, as often observed in \texttt{gpt-4o} and \texttt{o3-mini}. These models often skip the last evaluation of the line \code{while condition:} before exiting the loop, without following the explicit step-by-step demonstrations in the context.

\section{Related Works}

Following the framework introduced by~\citet{morris2024levelsagioperationalizingprogress}, \benchname~evaluates the ability to execute unambiguously defined simple routines step-by-step without deviation, a ``compiler''-like ``level-0'' intelligence that complex reasoning processes rely upon. The ability to maintain correctness by reliably executing simple rules also echoes the precise reasoning framework proposed by~\citet{shalevshwartz2024artificialexpertintelligencepacreasoning}, who formalize the compounding errors in multi-step problem solving. While \benchname~focuses on the coding domain, the task setup is distinct from existing works on code generation or completion~\citep{zhuo2024bigcodebench, jain2025livecodebench}. Instead, models are graded on their ability to generate concrete, often long, execution traces, which are proven to benefit code reasoning tasks~\citep{ni2024nextteachinglargelanguage}. 

\nparagraph{Teaching models to execute.} Prior works often explicitly \emph{teach} models to generate execution trace~\citep{liu-etal-2023-code}. For instance, ~\citet{zaremba2014learning} develop curriculum training methods to teach models to execute short programs, such as addition and memorization. Other works~\citep{NEURIPS2020_c8b9abff,Veličković2020Neural} train neural LMs to execute subroutines, such as sorting or classical graph algorithms, by developing novel architectures or training strategies. More recently, ~\citet{markeeva2024clrstextalgorithmicreasoninglanguage} train LMs to produce traces for a set of algorithms from the textbook \emph{Introduction to Algorithms}, and demonstrate significantly better performance than general-purpose pre-trained large language models. Recent works also elicit the execution process during test time by prompting large pre-trained language models. The reasoning process over a scratchpad~\citep{nye2021show}, however, is often a mixture of natural and formal languages~\citep{chae-etal-2024-language,arakelyan2024flare,zhou2022teachingalgorithmicreasoningincontext}, which inhibit proper evaluation of procedural correctness.

\nparagraph{Evaluation of procedural correctness.} Existing reasoning benchmarks, such as MATH~\citep{hendrycks2021measuring}, evaluate exclusively the correctness of \emph{outcomes} 
 (e.g., the solution inside \texttt{\$$\backslash$boxed\{*\}\$}).
We argue that a ``reasoning'' model should produce correct reasoning processes, which necessarily lead to correct solutions. 
Program execution is a natural subject to study given its clarity, verifiability, and constrained rule-based nature. Prior works that challenge models with code execution tasks~\citep{tufano2023predicting,pmlr-v235-gu24c,la2024code} are often limited by the data curation process or the fixed pre-defined set of algorithms.
Recent works also use natural language to define problems with unique and deterministic reference trajectory. ~\citet{gui2024logicgamebenchmarkingrulebasedreasoning} define rule-based games or puzzles~\citep{lin2025zebralogicscalinglimitsllms}  with varying difficulty level to evaluate both the ability to plan and execute rules; ~\citet{fujisawa2024procbenchbenchmarkmultistepreasoning} and ~\citet{ye2025longprocbenchmarkinglongcontextlanguage} both use natural languages to describe constraints and expected routines in the context, covering synthetic tasks (e.g., toy string manipulation) and tasks closer to real scenarios (e.g., travel planning). In contrast with these works on following natural language routines~\citep{chen-etal-2024-sifo}, \benchname~provides means to scalable generation of program-oriented instances, a more general framework admitting any problems that can be framed as programs. Additional related works and discussion on future directions are provided in Appendix~\ref{sec:app_add_disc}.

\section{Conclusion}

We presented \benchname, a synthetic language model benchmark for testing a special form of deduction, or the ability to reliably execute simple rules throughout predefined multi-step procedures. Specifically, given synthetic programs, \benchname~scores models based on their ability to generate accurate, step-by-step execution traces without error. 
The program-oriented synthetic benchmark enables scalable generation of test instances and allows for precise evaluation of procedural correctness, an aspect often overlooked by benchmarks focusing on outcome correctness. 
We evaluated $20$ recent language models on a base evaluation set of \benchname. Results revealed key limitations in maintaining correctness during simple multi-step procedures, even among the latest reasoning-enhanced models. Additionally, we employed \benchname~as a constrained environment to explore test-time scaling and other task variations, such as a more difficult transduction setting. 
Our findings suggest several directions for future research, including systematic investigation of various test-time scaling axes, extending \benchname~to evaluate reasoning ability beyond ``level-0,'' and leveraging scalable synthetic data generation to build more reliable reasoning systems.
\section*{Acknowledgments}
We would like to thank Yejin Choi for the helpful discussion.

\bibliography{colm2025_conference}
\bibliographystyle{colm2025_conference}

\appendix
\newpage
\section{Grammar for generating synthetic programs} \label{sec:appendix_grammar}

\begin{figure}[!h]
\begin{tcolorbox}[title=Baby Python Grammar Production Rules]
\vspace{0.2em}
$\langle program\rangle  ::=  \langle stmt\_lst\rangle  $ \\ 
$\langle stmt\_lst\rangle  ::= \langle stmt\rangle  \mid \langle stmt\rangle  \langle stmt\_lst\rangle $ \\ 
$\langle stmt\rangle  ::= \langle assignment\rangle  \mid \langle if\_block\rangle  \mid \langle while\_block\rangle  \mid \langle list\_op\rangle $ \\
\\
\textbf{\# assignment} \\
$\langle assignment\rangle  ::= \langle num\_assignment\rangle  \mid \langle bool\_assignment\rangle $ \\
$\langle num\_assignment\rangle  ::=  \langle var\rangle  \text{ = } \langle expr\rangle $ \\
$\langle bool\_assignment\rangle  ::= \langle bool\_var\rangle   \text{ = } \langle bool\_expr\rangle $ \\

\textbf{\# if block} \\
$\langle if\_block\rangle  ::= \text{if } \langle bool\_var\rangle  :  \langle stmt\_lst\rangle $ \\
 \\
\textbf{\# while block} \\
$\langle init\_cnter\rangle  ::= \text{cnter}  \text{ = 0}     $ \\
$\langle increment\_cnter\rangle  ::= \text{cnter} \text{ = } \text{cnter} + \langle cnter\_increment\_number\rangle  $ \\
$\langle while\_cond\rangle  ::= \langle while\_cond\_var\rangle  \text{ = } \text{cnter !=} \langle while\_cond\_number\rangle  $ \\
$\langle while\_block\rangle  ::= \langle while\_block\_nh\rangle  \mid \langle init\_cnter\rangle  \langle while\_cond\rangle  \text{ while } \\
\langle while\_cond\_var,*\rangle  :  \langle stmt\_lst\rangle  \langle increment\_cnter\rangle  \langle while\_cond,*\rangle $ \\
$\langle while\_block\_nh\rangle  ::= \text{ while True} :   \langle stmt\_lst\rangle $ \\
 \\
\textbf{\# list op} \\
$\langle list\_op\rangle  ::= \langle list\_var\rangle  \text{.append(} \langle operand\rangle  \text{)} \mid \langle list\_var\rangle  \text{.pop()}$ \\
\\
\textbf{\# expr} \\
$\langle operand\rangle  ::= \langle number\rangle  \mid \langle var\rangle $ \\
$\langle expr\rangle  ::= \langle operand\rangle  \mid \langle arithm\_expr\rangle  \mid \langle list\_var\rangle  [ \langle number \rangle  ]$ \\
$\langle arithm\_expr\rangle  ::= \langle operand\rangle  + \langle operand\rangle \mid \langle operand\rangle  - \langle operand\rangle $ \\
$\langle bool\_expr\rangle  ::= \langle operand\rangle  \text{ == } \langle operand\rangle \mid \langle operand\rangle  \text{ != } \langle operand\rangle$ \\
\end{tcolorbox}
\end{figure}

We design a ``baby'' version of Python by enforcing multiple constraints. Specifically:
\begin{itemize}[leftmargin=*]
    \item The generated program is a standalone Python function. The above production rules delineate the generation of the function body. We prepend, e.g., \code{define function(x, lst\_y, cond\_z):} before the function body and replace the input arguments \code{x, lst\_y, cond\_z} with randomly sampled values, such as \code{function(x=7, lst\_y=[1,2,3], cond\_z=False)}, when calling the function. 
    \item The variables are one of the following: integer, list of integers, and boolean value. Additionally, only list indexing is allowed to access a list (i.e., no slicing). The operations to a list can be either append or pop.  We restrict the variable type, data structure and associated operations to enforce simplicity of the program and the amount of knowledge prior required to execute the program. 
    \item Supported binary operations include only \code{+}, \code{-}, \code{==}, and \code{!=}, as many models struggle to reliably execute other operations such as \code{*}, \code{\%}, and \code{<}. 
    \item Complex single-line expressions are disabled to ensure the simplicity of transition between lines. For instance, we do not allow evaluation of long expressions, such as \code{(x + lst[5] - y - 6)}. Preliminary experiments reveal that models sometimes fail to correctly parse the conditions such as \code{if x != 0:}. Therefore in the production rules, we first evaluate the condition expression, and then assign the value to a boolean variable (e.g., \code{cond\_x}). The condition variable is then used for \emph{if} and \emph{while} blocks to separate condition evaluation from branching or jump. 
    \item In our base evaluation set, we exclude \code{else:} because this line is occasionally skipped by the models, or by the tracers depending on low-level compiler optimization. This complicates the evaluation process, specifically aligning model responses with ground-truths, therefore we opt to use solely the \code{if condition:} to enable conditional branching.  
    \item To enforce terminable \emph{while} loops, and to reduce execution errors, we impose context-dependent rules to the \emph{while} statement. Concretely, the terminable \emph{while} block always starts with the initialization of a \emph{counter} variable that controls the number of looping steps. The termination condition is evaluated both before the line of \code{while cond:} and at the end of the \emph{while} block. The \texttt{<while cond, *>} in the production rules denote a context-sensitive expansion, forcing the production to use the previously generated variable name for \code{while cond:}.
    \item Terminals are randomly selected from a pool of names given their types. In our current configuration, variables are named with a single letter, list variables are named as \texttt{lst\_{x}} where \texttt{x} is a single letter, boolean conditions are named following the pattern \texttt{cond\_{x}} where \texttt{x} is a single letter. We provide interface to override the naming of variables and certain production rules for flexibly configuring \benchname.
\end{itemize}

Various configurations or filters can be applied during synthetic program generation. In our main results, we adopt the configurations below:
\begin{itemize}
    \item Input integers are capped at 10
    \item Input list sizes are randomly sampled between 5 and 10.
    \item The maximum input program lines of code is 50
    \item The maximum scope depth is 1, i.e., there is no nested loops or conditional branches. 
    \item The \emph{while} loops terminate when the counter reaches a number $\leq$ 100.
    \item The maximum expansion depth of production rules is set to 200.
    \item We generate 64 few-shot step-by-step demonstrations for each program.
\end{itemize}
The above configurations can be modified to generate more difficult \benchname~split, e.g., increasing the size of input lists, the maximum input integers, enabling deep nested loops or branches, filtering programs and inputs that require large number of trace steps to complete. 

\section{Ablation of simple operations \& procedures} \label{sec:appendix_ablate_prior}

The synthetic programs are combinations of basic programming elements, which we ablate in this section. Concretely, we generate programs that contain only one of the basic ingredients of conventional programming languages by overriding existing production rules, such as setting \code{<stmt> ::= <assignment> | <if\_block> } to test the conditional branching control. For the ablation studies, we limit the program's lines of code to be at maximum 15, and test traces much shorter than even the short split in our default data except for the ablation of \emph{while} blocks. The ablations serve as a ``prior check'' of the ability to understand basic programming routines. 

Table~\ref{tab:ablate_prior} shows that even small models of 7B parameters can perform the evaluation of arithmetic (max integer values set to 10) and comparison expressions with high pass rate, when the operands are not variables. When the operands are previously assigned variable names e.g. \code{z = y + x}, (\textcolor{violet}{\texttt{w/ var. look up}} in the table), models show consistent degradation, with the 7B models degrading more. As such, during program generation we exclude the statements where both operands are variable names. To test the control flow, we insert simple assignment statements inside the \emph{if} block or \emph{while} loop. All models follow the control flow with acceptable accuracy though non-marginal gaps remain between small and larger models. The lists can be considered as the ``memory'' that a processor operates on. We independently test list indexing, append and pop operations at three list sizes $L$: short ($L \in [5, 10]$), medium ($L \in [25, 30]$), and long ($L \in [50, 55]$). Results demonstrate significantly worse performance for longer lists, even for the closed-source model \texttt{gpt-4o}. Additionally, list indexing with a previously assigned variable (e.g., \code{list[x]}, \textcolor{violet}{\texttt{list w/ var. look up}} in the table) also leads to consistent degradation, therefore, the list indexing operation in our baseline evaluation set always uses numbers instead of variable names as list indices. 

\begin{table}[]
\centering
\scalebox{0.75}{
\begin{tabular}{@{}lc|ccccc@{}}
\toprule
 & \textbf{\begin{tabular}[c]{@{}c@{}}\# Target \\Steps\end{tabular}}  & \textbf{\begin{tabular}[c]{@{}l@{}}gpt-4o\end{tabular}} & \textbf{\begin{tabular}[c]{@{}c@{}}qwen2.5 \\ coder-7b-inst\end{tabular}} & \textbf{\begin{tabular}[c]{@{}c@{}}qwen2.5-7b \\ inst-1m\end{tabular}} & \textbf{\begin{tabular}[c]{@{}c@{}}qwen2.5-coder \\ -32b-inst\end{tabular}} & \textbf{\begin{tabular}[c]{@{}c@{}}qwen2.5 \\ -32b-inst\end{tabular}} \\ \midrule
\textbf{Arithm. and comp op.}& & & & & & \\
\hspace{1em}+ and - & 5.3& 5.3& 5.2& 5.3 & 5.2 & 5.3 \\
\hspace{1em}+ and -, w/ var. look up & 5.3& 5.0& 4.3& 4.2 & 4.5 & 4.7 \\
\hspace{1em}== and !=& 5.3& 5.3& 5.2& 5.3 & 5.2 & 5.1 \\
\hspace{1em}== and != w/ var. look up& 5.3& 5.3& 4.1& 3.3 & 4.8 & 5.1 \\ \midrule
\textbf{Control} & & & & & & \\
\hspace{1em} if block & 4.51 & 4.51 & 4.34 & 3.97 & 4.47 & 4.48 \\ 
\hspace{1em} while block & 12.0 & 11.9 & 11.6 & 11.2& 11.7& 11.9\\ \midrule
\textbf{Memory (list) op \& access}& & & & & & \\
\hspace{1em}\texttt{x = lst[*]} & & & & & & \\
\hspace{2em} short list & 5.3& 5.2& 2.4& 1.4 & 4.7 & 5.0 \\
\hspace{2em} short list w/ var. look up& 5.4& 4.6& 1.5& 1.3 & 2.9 & 3.5 \\
\hspace{2em} medium list & 5.3& 5.0& 0.5& 0.3 & 1.2 & 2.4 \\
\hspace{2em} medium list w/ var. look up& 5.3& 3.5& 0.4& 0.5 & 1.3 & 1.9 \\
\hspace{2em} long list & 5.3& 3.2& 0.3& 0.2 & 0.5 & 0.7 \\
\hspace{2em} long list w/ var. look up& 5.3& 2.8& 0.4& 0.5 & 0.8 & 1.5 \\
\hspace{1em}\texttt{.append(*)} & & & & & & \\
\hspace{2em} short list & 5.3& 5.3& 4.2& 4.5 & 5.1 & 5.3 \\
\hspace{2em} medium list & 5.3& 5.3& 2.9& 3.7 & 4.8 & 5.0 \\
\hspace{2em} long list & 5.4& 5.3& 3.0& 3.6 & 5.0 & 5.2 \\
\hspace{1em}\texttt{.pop()} & & & & & & \\
\hspace{2em} short list & 5.3& 5.3& 4.0& 2.8 & 4.8 & 4.6 \\
\hspace{2em} medium list & 5.3& 5.1& 1.4& 0.5 & 2.0 & 3.6 \\
\hspace{2em} long list & 5.4& 5.3& 0.6& 0.5 & 2.2 & 3.3 \\
\bottomrule
\end{tabular}}
\caption{Average number of steps to the first error in ablation study described in Appendix~\ref{sec:appendix_ablate_prior}. }
\label{tab:ablate_prior}
\end{table}

\section{Additional Discussion \& Future Directions} \label{sec:app_add_disc}

\nparagraph{Additional Related Works.} \benchname~focuses on the coding domain~\citep{pmlr-v202-gao23f,chen2023program}. We select Python for its better readability and its prevalence in training data. Additionally, programs can be executed to produce traces as verifiable intermediate steps. This work tests models' ability to produce execution trace or simulating the execution process via generation~\citep{chen2024reasoning,li2024chaincodereasoninglanguage,la2024code}, instead of calling external symbolic reasoner~\citep{pan-etal-2023-logic}. The procedures are deterministically defined, thus separating the deduction (or applying general rules to instances) from searching over multiple valid deduction steps~\citep{saparov2023testing,saparov2023language}. Moreover, similar to other synthetic reasoning benchmarks~\citep{mirzadeh2025gsmsymbolic,zhou2025gsminfinitellmsbehaveinfinitely}, programs in \benchname~can be conveniently generated following rule-based methods. While the subject is in programming language, the task setup is different than code completion tasks~\citep{ni-etal-2024-l2ceval,liu2024repobench} or other related coding-specific tasks~\citep{yan-etal-2024-codescope}. 

\nparagraph{Limited Domain.} While \benchname~offers a systematic evaluation of procedural correctness, it is limited by its specific domain in programming language. Despite the imposed constraints for creating a more general and less coding-specific setup, it remains unclear to what extent the performance on step-by-step execution of simple Python procedures generalizes to realistic, more complex, long-horizon agentic tasks. Fully aware of such a limitation, we argue that \benchname~nonetheless provides a valuable check of precise and consistent rule-following, thus providing a more fine-grained diagnosis of reasoning capability, complementing existing advanced reasoning benchmarks that often test capabilities in collections. We leave systematic investigation on the correlation between model performance on \benchname~and other real-world tasks as future work. 

\nparagraph{Future Directions.} Besides stress testing models with \benchname~data generated with harder configurations, multiple future directions are promising to explore which are out of the scope of this work:
\begin{itemize}[leftmargin=*]
    \item \textbf{Relationship with real-world tasks}: As mentioned above, a promising direction is to explore how \benchname~can serve as a diagnostic test for real-world~\citep{jimenez2024swebench} agentic tasks. To understand this, one can perform correlational study between a variety of real-world multi-step tasks and instantiations of \benchname~with various configurations. Tasks covering diverse domains (e.g., administrative work flow, tabular data manipulation, etc) and capabilities (e.g., induction, abduction) can help better reveal relationship between \benchname~and realistic scenarios. Due to the absence of enough suitable real-world tasks and the difficulty in evaluating them, we leave this direction as future work.  
    \item \textbf{Systematic study of various test-time scaling dimensions}: We have shown the individual and synergistic gains of three test-time scaling dimensions: input context length, number of solutions for majority voting, and long chain-of-thought inference. However, the study remains superficial without probing deep into the trade-offs among these dimensions, which are likely model-specific. Systematically investigating the relationship between system specialty (e.g., domain specific, long-context, fast inference) vs. various test-time scaling dimensions can help inform better cost-effective model deployment.
    \item \textbf{Reasoning Effort vs. Task Difficulty}: Despite the simplicity of procedures in \benchname~(e.g., incrementing a counter, jumping to the right line based on the conditions), current reasoning models often sub-optimally spend thousands of tokens on ``self-reflective thinking.'' Ideally, reasoning effort increases with task difficulty. The examples in our baseline \benchname~are designed to be simple \emph{on purpose} (see Appendix~\ref{sec:appendix_grammar} and ~\ref{sec:appendix_ablate_prior} for constraints and ablation studies for justifying some of the simplifications), therefore much less reasoning effort is expected than what is observed in existing long-CoT models. While the number of thought tokens does reduce as the number of demonstrations in the context increases (Table~\ref{tab:num_thought_tokens}), more systematic study is required to fully understand the behavior of current reasoning-enhanced models. We list a few methods to increase task difficulty of \benchname: increase input list size, increase max scope depth (to enable nested loops), enable long expression evaluation, enable variable look-ups, increase program length, enable other arithmetic operations, etc. Effectively reducing reasoning verbosity could help reduce latency, on top of faster inference algorithms or architectural modifications. 
    \item \textbf{Teach models to follow procedures}: \benchname~provides scripts to generate large amount of synthetic data according to specified configurations. This allows for fine-tuning experiments on scalable generation of synthetic data. The controllable framework can potentially help investigate the effects of SFT and RL methods on narrowing the gap between \emph{pass@1} and \emph{pass@k}, and how these training strategies differ from each other in terms of generalization.
    \item \textbf{Architectural limitations}: Examples in \benchname~are compositions of fundamental operations that may help reveal key architectural constraints. The ``atomic'' operations, such as accurately copying lists followed by simple modifications (pop or append), or correctly retrieving and applying conditions from distant code sections (while loop), can expose fundamental limitations of existing architectures and help inform future improvements.
\end{itemize}

\section{Prompt template \& Example} \label{sec:prompt_example}

\begin{figure}[!h]
\begin{tcolorbox}
Please execute the following program by outputting a trace of the program execution. The trace should include the line number, the variable name and the updated value of the variable after executing that line. Please follow the examples below: \\
\\
Program:\\
\text{```} \\
\textcolor{gray}{\{input\_program\}} \\
\text{```} \\
\textcolor{gray}{\{fewshots\}}\\
\\
Input:\\
\text{```} \\
function(\textcolor{gray}{\{function\_args\}}) \\
\text{```} \\
\\
Please follow the format above and provide the program execution trace starting with \text{```}L2, \\
\\
Output:
\end{tcolorbox}
\caption{Template for for the default setup of \benchname. The input programs, few-shot demonstrations, and test input arguments are filled on the fly during evaluation.}
\label{fig:prompt_template}
\end{figure}

\begin{table}[!t]
    \centering
    \scriptsize
    \begin{tabular}{p{0.99\linewidth}}
    \toprule
Please execute the following program by outputting a trace of the program execution. The trace should include the line number, the variable name and the updated value of the variable after executing that line. Please follow the examples below:\\
\\
Program:\\
```\\
L1 def function(y, v, w, lst\_x, lst\_z, lst\_w, cond\_y, cond\_x):\\
L2 \hspace{1em} if cond\_y: \\
L3 	\hspace{2em} lst\_w.append(y)\\
L4 \hspace{1em} lst\_x.pop()\\
L5 \hspace{1em} 	lst\_x.append(8)\\
L6 	\hspace{1em} cond\_y = 6 == 3\\
L7 \hspace{1em} 	if cond\_y: \\
L8 	\hspace{2em} 	cond\_c = 6 == 1\\
L9 	\hspace{2em} 	lst\_z.pop()\\
L10 \hspace{1.5em} 	lst\_z.pop()\\
L11 \hspace{0.5em}	lst\_z.append(w)\\
L12 \hspace{0.5em}	if cond\_x: \\
L13 \hspace{1.5em}		lst\_z.append(3)\\
L14 \hspace{1.5em}		cond\_z = 5 != 0\\
L15 \hspace{1.5em}		i = 3 + 9\\
L16 \hspace{1.5em}		lst\_w.append(w)\\
L17 \hspace{0.5em}	lst\_x.pop()\\
L18 \hspace{0.5em}	lst\_w.append(v)\\
L19 \hspace{0.5em}	cond\_d = 1 != v\\
L20 \hspace{0.5em}	if cond\_y: \\
L21 \hspace{1.5em}		lst\_x.pop()\\
L22 \hspace{0.5em}	return\\
```\\
\\
Input:\\
```\\
function(y=0, v=2, w=8, lst\_x=[9, 3, 9, 9, 7, 8], lst\_z=[6, 6, 5, 6, 4, 7, 2, 8, 1], lst\_w=[0, 2, 6, 8, 1], cond\_y=False, cond\_x=True)\\
```\\
\\
Output:\\
```\\
L2,\\
L4,lst\_x:[9,3,9,9,7]\\
L5,lst\_x:[9,3,9,9,7,8]\\
L6,cond\_y:False\\
L7,\\
L11,lst\_z:[6,6,5,6,4,7,2,8,1,8]\\
L12,\\
L13,lst\_z:[6,6,5,6,4,7,2,8,1,8,3]\\
L14,cond\_z:True\\
L15,i:12\\
L16,lst\_w:[0,2,6,8,1,8]\\
L17,lst\_x:[9,3,9,9,7]\\
L18,lst\_w:[0,2,6,8,1,8,2]\\
L19,cond\_d:True\\
L20,\\
L22,\\
```\\
\\
Input:\\
```\\
function(y=8, v=2, w=7, lst\_x=[9, 3, 5, 2, 6, 0], lst\_z=[0, 8, 4, 5, 8, 4, 4], lst\_w=[2, 8, 2, 1, 7, 9, 9, 5, 8, 5], cond\_y=False, cond\_x=False)\\
```\\
\\
Please follow the format above and provide the program execution trace starting with ```L2,\\
\\
Output:\\
\bottomrule
    \end{tabular}
    \caption{Example prompt in \benchname. Only one-shot demonstration is provided in this example. The number of few-shot demonstrations is an adjustable hyperparameter in \benchname. We set to 4 in our main experiments.}
    \label{tab:example_prompt}
\end{table}
See Figure~\ref{fig:prompt_template} and Table~\ref{tab:example_prompt}.

\section{Complete results} \label{sec:appendix_complete_results}

\begin{table}[ht!]
\centering
\resizebox{\textwidth}{!}{%
\begin{tabular}{lccccccc}
\toprule
Model                         & pass@1 & pass@5 & pass@15 & pass@31 & maj@5 & maj@15 & maj@31 \\ \midrule
\multicolumn{8}{c}{\textit{\textbf{Short Traces}}}                                                              \\ \midrule
o1 & 96.60 & - & - & - & - & - & -\\
o3-mini & 98.80 & - & - & - & - & - & -\\
gpt-4o-2024-11-20 & 85.80 & - & - & - & - & - & -\\
claude-3-7-sonnet & 98.60 & - & - & - & - & - & -\\ 
claude-3-5-sonnet-20241022 & 89.20 & - & - & - & - & - & -\\
QwQ-32B                       & 95.26  & 99.70  & 99.90   & 100.00  & 98.60    
   & 98.60        & 99.20       \\
DeepSeek-R1-Distill-Llama-70B & 54.27  & 83.30  & 92.50   & 92.50   & 64.00       & 70.80        & 73.40        \\
DeepSeek-R1-Distill-Qwen-32B  & 73.21  & 98.00  & 100.00  & 100.00  & 90.00       & 94.40        & 95.60        \\
DeepSeek-R1-Distill-Qwen-14B  & 45.34  & 87.00  & 98.00   & 99.00   & 63.00       & 77.20        & 81.60        \\
DeepSeek-R1-Distill-Qwen-7B   & 10.72  & 37.00  & 59.00   & 71.00   & 14.80       & 25.80        & 29.80        \\
Deepseek-R1                   & 98.10  & 100.00 & 100.00  & 100.00  & 99.80       & 99.60        & 100.00       \\
Llama-3.1-405B-Instruct       & 77.20  & 93.40  & 97.00   & 98.20   & 81.00       & 84.60        & 85.60        \\
Llama-3.1-70B-Instruct        & 58.19  & 83.20  & 90.60   & 92.90   & 67.60       & 70.40        & 70.60        \\
Llama-3.1-8B-Instruct         & 7.24   & 19.00  & 30.00   & 38.00   & 8.60        & 12.00        & 13.00        \\
Qwen2.5-72B-Instruct          & 65.79  & 88.40  & 93.10   & 94.90   & 76.60       & 79.20        & 80.80        \\
Qwen2.5-32B-Instruct          & 61.96  & 87.10  & 93.30   & 95.20   & 72.80       & 76.40        & 77.20        \\
Qwen2.5-14B-Instruct          & 30.00  & 58.00  & 72.90   & 80.50   & 38.40       & 41.20        & 42.60        \\
Qwen2.5-7B-Instruct           & 20.80  & 47.00  & 62.00   & 70.00   & 28.20       & 34.40        & 36.80        \\
Qwen2.5-Coder-32B-Instruct    & 57.94  & 83.20  & 91.00   & 93.90   & 66.60       & 69.80        & 71.40        \\
Qwen2.5-Coder-7B-Instruct     & 28.00  & 59.00  & 74.00   & 81.00   & 36.80       & 42.20        & 47.80        \\ \midrule
\multicolumn{8}{c}{\textit{\textbf{Medium Traces}}}                                                             \\ \midrule
o1 & 97.40 & - & - & - & - & - & -\\
o3-mini & 67.20 & - & - & - & - & - & -\\
gpt-4o-2024-11-20 & 76.40 & - & - & - & - & - & -\\
claude-3-7-sonnet & 97.00 & - & - & - & - & - & -\\ 
claude-3-5-sonnet-20241022 & 87.80 & - & - & - & - & - & -\\
QwQ-32B                       & 84.67  & 99.40  & 100.00   & 100.00  & 93.00   
   & 94.80        & 96.00       \\
DeepSeek-R1-Distill-Llama-70B & 72.44  & 93.90  & 98.00   & 98.80   & 80.20       & 87.40        & 88.60        \\
DeepSeek-R1-Distill-Qwen-32B  & 86.49  & 99.00  & 100.00  & 100.00  & 95.60       & 97.60        & 98.00        \\
DeepSeek-R1-Distill-Qwen-14B  & 66.74  & 94.00  & 99.00   & 100.00  & 78.80       & 87.00        & 90.00        \\
DeepSeek-R1-Distill-Qwen-7B   & 4.33   & 16.00  & 30.00   & 41.00   & 5.20        & 12.20        & 14.40        \\
Deepseek-R1                   & 95.88  & 99.80  & 99.80   & 99.80   & 99.20       & 99.40        & 99.80        \\
Llama-3.1-405B-Instruct       & 61.70  & 84.40  & 91.20   & 93.90   & 68.20       & 69.80        & 71.60        \\
Llama-3.1-70B-Instruct        & 40.06  & 67.40  & 79.40   & 85.20   & 47.80       & 51.00        & 51.80        \\
Llama-3.1-8B-Instruct         & 7.29   & 18.10  & 27.70   & 35.30   & 7.60        & 12.00        & 13.20        \\
Qwen2.5-72B-Instruct          & 56.99  & 81.40  & 88.80   & 91.90   & 63.20       & 67.80        & 69.60        \\
Qwen2.5-32B-Instruct          & 49.61  & 77.70  & 87.00   & 90.90   & 60.00       & 64.60        & 65.60        \\
Qwen2.5-14B-Instruct          & 22.34  & 48.70  & 64.80   & 73.30   & 28.20       & 30.80        & 32.80        \\
Qwen2.5-7B-Instruct           & 9.60   & 28.00  & 42.00   & 52.00   & 13.00       & 19.00        & 21.80        \\
Qwen2.5-Coder-32B-Instruct    & 45.48  & 71.20  & 81.20   & 85.30   & 51.60       & 56.00        & 57.80        \\
Qwen2.5-Coder-7B-Instruct     & 16.74  & 38.00  & 54.00   & 63.00   & 21.60       & 28.40        & 31.80        \\ \bottomrule
\end{tabular}%
}
\caption{Detailed results for Table~\ref{tab:main_res_tab}.}
\label{tab:main_big_0}
\end{table}

\begin{table}[ht!]
\centering
\resizebox{\textwidth}{!}{%
\begin{tabular}{lccccccc}
\toprule
Model                         & pass@1 & pass@5 & pass@15 & pass@31 & maj@5 & maj@15 & maj@31 \\ \midrule
\multicolumn{8}{c}{\textit{\textbf{Long Traces}}}                                                               \\ \midrule
o1 & 91.40 & - & - & - & - & - & -\\
o3-mini & 46.20 & - & - & - & - & - & -\\
gpt-4o-2024-11-20 & 74.20 & - & - & - & - & - & -\\
claude-3-7-sonnet & 97.00 & - & - & - & - & - & -\\ 
claude-3-5-sonnet-20241022 & 82.20 & - & - & - & - & - & -\\
QwQ-32B                       & 84.47  & 98.70  & 99.70   & 100.00  & 92.60   
   & 95.40        & 96.20       \\
DeepSeek-R1-Distill-Llama-70B & 65.36  & 90.80  & 96.50   & 98.00   & 77.20       & 81.80        & 83.40        \\
DeepSeek-R1-Distill-Qwen-32B  & 83.58  & 98.00  & 100.00  & 100.00  & 92.00       & 94.80        & 94.80        \\
DeepSeek-R1-Distill-Qwen-14B  & 60.96  & 90.00  & 96.00   & 98.00   & 75.20       & 85.60        & 85.80        \\
DeepSeek-R1-Distill-Qwen-7B   & 0.19   & 1.00   & 2.00    & 3.00    & 0.40        & 0.60         & 0.60         \\
Deepseek-R1                   & 94.84  & 99.30  & 99.70   & 99.80   & 99.20       & 99.20        & 99.60        \\
Llama-3.1-405B-Instruct       & 59.20  & 81.60  & 89.10   & 92.50   & 63.60       & 66.20        & 68.00        \\
Llama-3.1-70B-Instruct        & 36.58  & 66.20  & 80.40   & 86.20   & 41.60       & 46.80        & 48.40        \\
Llama-3.1-8B-Instruct         & 5.66   & 15.50  & 24.90   & 31.30   & 6.60        & 8.80         & 9.20         \\
Qwen2.5-72B-Instruct          & 52.57  & 78.20  & 88.00   & 91.70   & 58.80       & 61.20        & 62.20        \\
Qwen2.5-32B-Instruct          & 47.62  & 75.00  & 86.00   & 90.00   & 54.40       & 60.40        & 62.80        \\
Qwen2.5-14B-Instruct          & 23.39  & 50.50  & 66.30   & 74.70   & 28.80       & 34.20        & 36.40        \\
Qwen2.5-7B-Instruct           & 8.59   & 22.80  & 36.10   & 18.20   & 11.60       & 16.00        & 18.20        \\
Qwen2.5-Coder-32B-Instruct    & 40.06  & 68.00  & 79.90   & 84.80   & 48.20       & 51.00        & 53.00        \\
Qwen2.5-Coder-7B-Instruct     & 13.29  & 34.00  & 49.00   & 59.00   & 17.80       & 24.60        & 26.40        \\ \midrule
\multicolumn{8}{c}{\textit{\textbf{Extra Long Traces}}}                                                          \\ \midrule
o1 & 82.60 & - & - & - & - & - & -\\
o3-mini & 30.00 & - & - & - & - & - & -\\
gpt-4o-2024-11-20 & 69.00 & - & - & - & - & - & -\\
claude-3-7-sonnet & 87.80 & - & - & - & - & - & -\\ 
claude-3-5-sonnet-20241022 & 73.80 & - & - & - & - & - & -\\
QwQ-32B                       & 81.99  & 95.70  & 98.30   & 99.20  & 88.40   
   & 92.20        & 93.00       \\

DeepSeek-R1-Distill-Llama-70B & 55.91  & 83.20  & 93.10   & 97.10   & 63.20       & 70.60        & 73.00        \\
DeepSeek-R1-Distill-Qwen-32B  & 78.23  & 95.00  & 98.00   & 99.00   & 87.60       & 90.80        & 91.20        \\
DeepSeek-R1-Distill-Qwen-14B  & 54.10  & 82.00  & 91.00   & 94.00   & 64.60       & 73.80        & 76.60        \\
DeepSeek-R1-Distill-Qwen-7B   & 0.00   & 0.00   & 0.00    & 0.00    & 0.00        & 0.00         & 0.00         \\
Deepseek-R1  & 77.04  & 88.10  & 90.80   & 92.10   & 85.40       & 88.80        & 90.00        \\
Llama-3.1-405B-Instruct       & 54.00  & 77.40  & 84.90   & 88.10   & 60.60       & 65.60        & 67.20        \\
Llama-3.1-70B-Instruct        & 34.57  & 62.40  & 74.30   & 79.40   & 42.80       & 47.80        & 48.00        \\
Llama-3.1-8B-Instruct         & 5.78   & 15.10  & 24.10   & 30.90   & 6.00        & 9.40         & 10.40        \\
Qwen2.5-72B-Instruct          & 45.60  & 68.70  & 77.30   & 81.10   & 52.00       & 54.20        & 54.80        \\
Qwen2.5-32B-Instruct          & 42.44  & 69.00  & 79.00   & 84.00   & 50.40       & 56.60        & 58.40        \\
Qwen2.5-14B-Instruct          & 19.11  & 42.70  & 57.40   & 66.20   & 24.20       & 28.20        & 30.80        \\
Qwen2.5-7B-Instruct           & 6.29   & 18.20  & 28.80   & 36.10   & 9.60        & 12.00        & 14.40        \\
Qwen2.5-Coder-32B-Instruct    & 32.38  & 57.00  & 68.10   & 73.60   & 39.80       & 45.00        & 44.80        \\
Qwen2.5-Coder-7B-Instruct     & 10.20  & 28.00  & 40.00   & 47.00   & 17.80       & 21.80        & 23.60        \\ \bottomrule
\end{tabular}%
}
\caption{Detailed results for Table~\ref{tab:main_res_tab}.}
\label{tab:main_big_1}
\end{table}
See Table~\ref{tab:main_big_0} and ~\ref{tab:main_big_1} for complete results.

\section{Additional Results} \label{sec:additional_res}
\begin{table}[]
\centering
\scalebox{0.9}{
\begin{tabular}{@{}lc|cc|ccc@{}} 
\toprule
 & \multicolumn{1}{l}{}& \multicolumn{2}{c}{\textbf{Steps To Err.}}& \multicolumn{3}{c}{\textbf{Trace Acc.}}\\ 
 & \begin{tabular}[c]{@{}c@{}}\# few\\ shot\end{tabular} & \begin{tabular}[c]{@{}c@{}}Single\\ Attempt\end{tabular} & \begin{tabular}[c]{@{}c@{}}maj vote\\ @ 31\end{tabular} & \begin{tabular}[c]{@{}c@{}}Single\\ Attempt\end{tabular} & \begin{tabular}[c]{@{}c@{}}maj vote\\ @ 31\end{tabular} & \begin{tabular}[c]{@{}c@{}}pass \\ @ 31\end{tabular} \\ \midrule
Short Trace& \multicolumn{1}{l}{}& \multicolumn{1}{l}{} & \multicolumn{1}{l}{}& \multicolumn{1}{l}{} & \multicolumn{1}{l}{}& \multicolumn{1}{l}{} \\
 & 2 & 8.3& 10.4 & 43.7& 66.0 & 94.6\\
 & 4 & 9.2& 11.1 & 53.5& 74.2 & 95.7\\
 & 8 & 9.6& 11.4 & 58.9& 77.6 & 98.1\\
 & 16& 10.0 & 11.6& 63.0& 81.4 & 97.8\\
 & 32& 10.0 & 11.4 & 62.5& 79.0 & 98.6\\ \midrule
Medium Trace & \multicolumn{1}{l}{}& \multicolumn{1}{l}{} & \multicolumn{1}{l}{}& \multicolumn{1}{l}{} & \multicolumn{1}{l}{}& \multicolumn{1}{l}{} \\
 & 2 & 65.4 & 73.4 & 65.5& 83.8 & 98.2\\
 & 4 & 69.9 & 76.0 & 72.9& 87.6 & 98.6\\
 & 8 & 71.3 & 76.7 & 75.1& 88.2 & 99.2\\
 & 16& 71.3 & 76.4 & 74.1& 86.6 & 98.6\\
 & 32& 67.9 & 72.7 & 69.5& 81.6 & 97.4 \\
 \bottomrule
\end{tabular}}
\caption{Scaling along all three test-time scaling dimensions. \texttt{R1-Distilled-Llama-70B}, a long-CoT model achieves further gains when provided with multiple step-by-step demonstrations and taking majority vote over solutions generated in parallel.}
\label{tab:r1_dt_llama}
\end{table}

\begin{table}
\centering
\small
\scalebox{0.9}{
\begin{tabular}{@{}lcccc@{}}
\toprule
\textbf{\# thought tokens} & \textbf{Short Trace}& \textbf{Medium Trace}& \textbf{Long Trace} & \textbf{Extra Long Trace}\\ \midrule
DeepSeek-R1-Distill-Qwen-7B& 1244& 3198& 4568 & 4366\\
DeepSeek-R1-Distill-Qwen-14B& 2618& 2294& 2084 & 1967\\
DeepSeek-R1-Distill-Qwen-32B& 1575& 1967& 1596 & 1753\\ 
QwQ-32B & 3360 & 7388 & 6793 & 6830 \\
\midrule
DeepSeek-R1-Distill-Llama-70B & \multicolumn{1}{l}{}& \multicolumn{1}{l}{}& \multicolumn{1}{l}{}& \multicolumn{1}{l}{} \\
\hspace{3em} 4-shot& 1820& 1279& 1176 & 1185\\
\hspace{3em} 8-shot& 1788& 1205& 1046 & 1105\\
\hspace{3em} 32-shot& 1290& 1187& 1134 & -\\ \bottomrule
\end{tabular}}
\caption{\small Thought tokens are enclosed by \texttt{<think>} and \texttt{</think>}. Within the same model series (R1-distilled), smaller models generate more thought tokens, with the 7B model frequently failing to exit the think pattern within the requested token budget. QwQ-32B produces significantly longer chain-of-thought than same-sized R1-distilled model.}
\label{tab:num_thought_tokens}
\end{table}
See Table~\ref{tab:r1_dt_llama} for test-time scaling of R1-Distill Llama and Table~\ref{tab:num_thought_tokens} for the number of thought tokens generated by reasoning-enhanced models.

\newpage
\section{Pass@k evaluation}
See Figure~\ref{fig:pass_k} for pass@k evaluation code.
\begin{figure}[!h]
\begin{tcolorbox}[]
\text{def pass\_at\_k(n, c, k):} \\
\text{\hspace{1em}} """ \\
\text{\hspace{1em}}     :param n: total number of samples \\
\text{\hspace{1em}}     :param c: number of correct samples \\
\text{\hspace{1em}}     :param k: k in pass@k \\
\text{\hspace{1em}}     """ \\
\text{\hspace{1em}}     if n - c $<$ k: \\
\text{\hspace{2em}}         return 1.0 \\
\text{\hspace{1em}}     return 1.0 - np.prod(1.0 - k / np.arange(n - c + 1, n + 1))\\
\end{tcolorbox}
\label{fig:pass_k}
\caption{Pass@K evaluation code}
\end{figure}

\section{Models}
\label{models}
See Table~\ref{tab:model_link} for a summary of models evaluated in the this work.
\begin{table}[h]
\centering
\resizebox{0.99\linewidth}{!}{
\begin{tabular}{@{}lccl@{}}
\toprule
Model Name & Size & Thinking & Huggingface / API \\
\midrule
o1~\citep{jaech2024openai} & - & Y & o1 \\
o3-mini~\citep{o3-mini} & - & Y & o3-mini \\
gpt-4o-2024-11-20~\citep{hurst2024gpt} & - & N & gpt-4o-2024-11-20 \\
claude-3-7-sonnet~\citep{claude37} & - & Y & claude-3-7-sonnet-20250219 \\
claude-3-5-sonnet-20241022~\citep{claude35} & - & N & claude-3-5-sonnet-20241022 \\
DeepSeek-R1~\citep{guo2025deepseek} & 37B / 671B & Y & deepseek-ai/DeepSeek-R1 \\
DeepSeek-R1-Distill-Qwen-32B~\citep{guo2025deepseek} & 32.5B & Y & deepseek-ai/DeepSeek-R1-Distill-Qwen-32B \\
DeepSeek-R1-Distill-Qwen-14B~\citep{guo2025deepseek} & 14.7B & Y & deepseek-ai/DeepSeek-R1-Distill-Qwen-14B \\
DeepSeek-R1-Distill-Qwen-7B~\citep{guo2025deepseek} & 7.6B & Y & deepseek-ai/DeepSeek-R1-Distill-Qwen-7B \\
DeepSeek-R1-Distill-Llama-70B~\citep{guo2025deepseek} & 70B & Y & deepseek-ai/DeepSeek-R1-Distill-Llama-70B \\
QwQ-32B~\citep{qwq32b} & 32.5B & Y & Qwen/QwQ-32B \\
Qwen2.5-72B-Instruct~\citep{qwen2.5} & 72.7B & N & Qwen/Qwen2.5-72B-Instruct \\
Qwen2.5-32B-Instruct~\citep{qwen2.5} & 32.5B & N & Qwen/Qwen2.5-32B-Instruct \\
Qwen2.5-14B-Instruct~\citep{qwen2.5} & 14.7B & N & Qwen/Qwen2.5-14B-Instruct \\
Qwen2.5-7B-Instruct~\citep{qwen2.5} & 7.6B & N & Qwen/Qwen2.5-7B-Instruct \\
Qwen2.5-Coder-32B-Instruct~\citep{qwen2.5} & 32.5B & N & Qwen/Qwen2.5-Coder-32B-Instruct \\
Qwen2.5-Coder-7B-Instruct~\citep{qwen2.5} & 7.6B & N & Qwen/Qwen2.5-Coder-7B-Instruct \\
Meta-Llama-3.1-405B-Instruct~\citep{grattafiori2024llama3herdmodels} & 405B & N & meta-llama/Llama-3.1-405B-Instruct \\
Meta-Llama-3.1-70B-Instruct~\citep{grattafiori2024llama3herdmodels} & 70B & N & meta-llama/Llama-3.1-70B-Instruct \\
Meta-Llama-3.1-8B-Instruct~\citep{grattafiori2024llama3herdmodels} & 8B & N & meta-llama/Llama-3.1-8B-Instruct \\
\bottomrule
\end{tabular}}
\caption{Model details.}
\label{tab:model_link}
\end{table}

\section{Example Long chain-of-thought output}
See Tables below for an example response by a top-performing open-source model \texttt{QwQ-32B}.
\begin{table}[]
    \centering
    \small
    \begin{tabular}{p{0.99\linewidth}}
    \toprule
    \textbf{QwQ-32B example output}  \\
    \midrule
    Okay, let's tackle this problem step by step. I need to trace the execution of the given Python function with the provided input. Let me start by understanding the function and the input parameters.\texttt{\textbackslash n}\texttt{\textbackslash n}The function is called function, and the input parameters are:\texttt{\textbackslash n}y=8, v=2, w=7,\texttt{\textbackslash n}lst\_x = [9, 3, 5, 2, 6, 0],\texttt{\textbackslash n}lst\_z = [0, 8, 4, 5, 8, 4, 4],\texttt{\textbackslash n}lst\_w = [2, 8, 2, 1, 7, 9, 9, 5, 8, 5],\texttt{\textbackslash n}cond\_y=False, cond\_x=False.\texttt{\textbackslash n}\texttt{\textbackslash n}First, I'll list out all the variables and their initial values before the function starts executing. The function's code is provided, so I'll go line by line.\texttt{\textbackslash n}\texttt{\textbackslash n}Starting with line L2: the first line is the function definition, but the actual execution starts when the function is called. The first executable line is L2's condition check.\texttt{\textbackslash n}\texttt{\textbackslash n}Let me go through each line step by step, keeping track of variable changes.\texttt{\textbackslash n}\texttt{\textbackslash n}Line L2: if cond\_y:\texttt{\textbackslash n}The current cond\_y is False (from input), so the condition is false. Therefore, the code inside the if block (line L3) is skipped. So after L2, nothing changes here. The trace entry for L2 is just a comma since no variables are updated.\texttt{\textbackslash n}\texttt{\textbackslash n}Next, line L4: lst\_x.pop()\texttt{\textbackslash n}The list lst\_x is [9,3,5,2,6,0]. The pop() method removes the last element by default. So popping removes 0, resulting in [9,3,5,2,6]. So after L4, lst\_x is now [9,3,5,2,6]. So the trace entry is L4,lst\_x: the new list.\texttt{\textbackslash n}\texttt{\textbackslash n}Line L5: lst\_x.append(8)\texttt{\textbackslash n}Appending 8 to the current lst\_x (after the pop), so the new list becomes [9,3,5,2,6,8]. So the trace entry is L5,lst\_x with the new list.\texttt{\textbackslash n}\texttt{\textbackslash n}Line L6: cond\_y = 6 == 3. The right side is a boolean expression. 6 ==3 is False, so cond\_y becomes False. The trace entry is L6,cond\_y:False.\texttt{\textbackslash n}\texttt{\textbackslash n}Line L7: if cond\_y: Since cond\_y is still False, the block under L7 (lines L8-L10) is skipped. So L7's trace is just a comma.\texttt{\textbackslash n}\texttt{\textbackslash n}Line L11: lst\_z.append(w). The current w is 7. The original lst\_z is [0,8,4,5,8,4,4]. Appending 7 gives [0,8,4,5,8,4,4,7]. So the trace entry is L11,lst\_z with the new list.\texttt{\textbackslash n}\texttt{\textbackslash n}Line L12: if cond\_x: cond\_x is False (from input), so the block (lines L13-L18) is skipped. Wait, wait. Wait, the lines under L12's if are L13 to L16, and then after that, L17 and L18 are outside the if block? Let me check the code structure again.\texttt{\textbackslash n}\texttt{\textbackslash n}Looking back at the program code:\texttt{\textbackslash n}\texttt{\textbackslash n}After L12 comes L13, which is indented under the if cond\_x block. Then L17 is after that block. So lines L13-L16 are inside the if cond\_x clause. Since cond\_x is False, those lines are skipped. So after L12, we proceed to L17.\texttt{\textbackslash n}\texttt{\textbackslash n}Wait, the code structure:\texttt{\textbackslash n}\texttt{\textbackslash n}L12: if cond\_x:\texttt{\textbackslash n}L13:    lst\_z.append(3)\texttt{\textbackslash n}L14:    cond\_z = 5 !=0\texttt{\textbackslash n}L15:    i =3+9\texttt{\textbackslash n}L16:    lst\_w.append(w)\texttt{\textbackslash n}Then L17 is outside the if block, so it's executed regardless.\texttt{\textbackslash n}\texttt{\textbackslash n}So after L12, since cond\_x is False, we skip L13-L16 and go to L17.\texttt{\textbackslash n}\texttt{\textbackslash n}Line L17: lst\_x.pop(). Current lst\_x after L5 was [9,3,5,2,6,8]. Wait, wait, let me retrace:\texttt{\textbackslash n}\texttt{\textbackslash n}Wait, after L4, lst\_x was [9,3,5,2,6], then L5 appends 8, making it [9,3,5,2,6,8]. Then L17 is next. So popping the last element (8) leaves [9,3,5,2,6]. So the trace entry for L17 is lst\_x: [9,3,5,2,6].\texttt{\textbackslash n}\texttt{\textbackslash n}Line L18: lst\_w.append(v). v is 2. The original lst\_w is [2,8,2,1,7,9,9,5,8,5]. Appending 2 gives [2,8,2,1,7,9,9,5,8,5,2]. So the trace entry is L18,lst\_w with the new list.\texttt{\textbackslash n}\texttt{\textbackslash n}Line L19: cond\_d = 1 != v. v is 2. 1 !=2 is True, so cond\_d is True. So the trace entry is L19,cond\_d:True.\texttt{\textbackslash n}\texttt{\textbackslash n}Line L20: if cond\_y: cond\_y is False (from L6), so the block (L21) is skipped. So L20's trace is a comma.\texttt{\textbackslash n}\texttt{\textbackslash n}Finally, return at L22, so the last entry is L22, with nothing else.\texttt{\textbackslash n}\texttt{\textbackslash n}Wait, let me check all steps again step by step to ensure I didn't miss anything.\texttt{\textbackslash n}\texttt{\textbackslash n}Starting over:\texttt{\textbackslash n}\texttt{\textbackslash n}Initial variables:\texttt{\textbackslash n}\texttt{\textbackslash n}y=8, v=2, w=7,\texttt{\textbackslash n}\texttt{\textbackslash n}lst\_x = [9,3,5,2,6,0]\texttt{\textbackslash n}\texttt{\textbackslash n}lst\_z = [0,8,4,5,8,4,4]\texttt{\textbackslash n}\texttt{\textbackslash n}lst\_w = [2,8,2,1,7,9,9,5,8,5]\texttt{\textbackslash n}\texttt{\textbackslash n}cond\_y = False, cond\_x = False.\texttt{\textbackslash n}\texttt{\textbackslash n}Execution steps:\texttt{\textbackslash n}\texttt{\textbackslash n}L2: if cond\_y (False) → skip L3.\texttt{\textbackslash n}\texttt{\textbackslash n}L4: lst\_x.pop() → removes last element (0) → [9,3,5,2,6]\texttt{\textbackslash n}\texttt{\textbackslash n}Trace: L4,lst\_x: [9,3,5,2,6]\texttt{\textbackslash n}\texttt{\textbackslash n}L5: append 8 → [9,3,5,2,6,8]\texttt{\textbackslash n}\texttt{\textbackslash n}Trace: L5,lst\_x: [9,3,5,2,6,8]\texttt{\textbackslash n}\texttt{\textbackslash n}L6: cond\_y = 6==3 → False → Trace: L6,cond\_y:False\texttt{\textbackslash n}\texttt{\textbackslash n}L7: if cond\_y (still False) → skip L8-L10.\texttt{\textbackslash n}\texttt{\textbackslash n}L11: append w (7) to lst\_z → original lst\_z was [0,8,4,5,8,4,4], so new is [0,8,4,5,8,4,4,7]\texttt{\textbackslash n}\texttt{\textbackslash n}Trace: L11,lst\_z: [0,8,4,5,8,4,4,7]\texttt{\textbackslash n}\texttt{\textbackslash n}L12: if cond\_x (False) → skip L13-L16.\texttt{\textbackslash n}\texttt{\textbackslash n}L17: pop lst\_x → current lst\_x is [9,3,5,2,6,8], so pop removes 8 → [9,3,5,2,6]\texttt{\textbackslash n}\texttt{\textbackslash n}Trace: L17,lst\_x: [9,3,5,2,6]\texttt{\textbackslash n}\texttt{\textbackslash n}L18: append v (2) to lst\_w → original lst\_w is [2,8,2,1,7,9,9,5,8,5], so new is adding 2 → [2,8,2,1,7,9,9,5,8,5,2]\texttt{\textbackslash n}\texttt{\textbackslash n}Wait, wait, wait! Wait, the initial lst\_w is [2,8,2,1,7,9,9,5,8,5]. So appending 2 gives the new list as that list plus 2 at the end. So the new list is [2,8,2,1,7,9,9,5,8,5,2]. So the trace entry for L18 is lst\_w with that list.\texttt{\textbackslash n}\texttt{\textbackslash n}L19: cond\_d = 1 != v → v is 2 → 1 !=2 → True → Trace: L19,cond\_d:True\texttt{\textbackslash n}\texttt{\textbackslash n}L20: if cond\_y (False) → skip L21.\texttt{\textbackslash n}\texttt{\textbackslash n}L22: return → so the last line is L22 with nothing \\
    \bottomrule
    \end{tabular}
    \label{tab:qwq_example_0}
\end{table}

\begin{table}[]
    \centering
    \small
    \begin{tabular}{p{0.99\linewidth}}
    \toprule
    \textbf{QwQ-32B example output (continued)} \\
    \midrule
    else.\texttt{\textbackslash n}\texttt{\textbackslash n}Now, let me check if I missed any lines or steps.\texttt{\textbackslash n}\texttt{\textbackslash n}Wait, after L12, we go to L17. Then L18, L19, L20, L22.\texttt{\textbackslash n}\texttt{\textbackslash n}Wait, after L12, the next line is L12's else? No, the code after L12 is L13 (indented under if cond\_x), but since cond\_x is false, we skip to after the block, which is L17.\texttt{\textbackslash n}\texttt{\textbackslash n}Yes, that's correct.\texttt{\textbackslash n}\texttt{\textbackslash n}Now, let me check each line's trace entries:\texttt{\textbackslash n}\texttt{\textbackslash n}The output should be:\texttt{\textbackslash n}\texttt{\textbackslash n}L2,\texttt{\textbackslash n}\texttt{\textbackslash n}L4,lst\_x:[9,3,5,2,6]\texttt{\textbackslash n}\texttt{\textbackslash n}L5,lst\_x:[9,3,5,2,6,8]\texttt{\textbackslash n}\texttt{\textbackslash n}L6,cond\_y:False\texttt{\textbackslash n}\texttt{\textbackslash n}L7,\texttt{\textbackslash n}\texttt{\textbackslash n}
    L11,lst\_z:[0,8,4,5,8,4,4,7]\texttt{\textbackslash n}\texttt{\textbackslash n}L12,\texttt{\textbackslash n}\texttt{\textbackslash n}L17,lst\_x:[9,3,5,2,6]\texttt{\textbackslash n}\texttt{\textbackslash n}L18,lst\_w:[2,8,2,1,7,9,9,5,8,5,2]
    \texttt{\textbackslash n}\texttt{\textbackslash n}L19,cond\_d:True\texttt{\textbackslash n}\texttt{\textbackslash n}L20,\texttt{\textbackslash n}\texttt{\textbackslash n}L22,\texttt{\textbackslash n}\texttt{\textbackslash n}Wait, but in the previous examples, sometimes there are more entries. Let me check if I missed any lines where variables are modified.\texttt{\textbackslash n}\texttt{\textbackslash n}Wait, let me go through each line again:\texttt{\textbackslash n}\texttt{\textbackslash n}Line L3: skipped because cond\_y was False initially.\texttt{\textbackslash n}\texttt{\textbackslash n}Line L4: executed, so lst\_x changes.\texttt{\textbackslash n}\texttt{\textbackslash n}Line L5: executed.\texttt{\textbackslash n}\texttt{\textbackslash n}Line L6: executed, cond\_y set to False.\texttt{\textbackslash n}\texttt{\textbackslash n}Line L7: condition is cond\_y (now False), so no execution of L8-10.\texttt{\textbackslash n}\texttt{\textbackslash n}Line L11: executed, appending w (7) to lst\_z.\texttt{\textbackslash n}\texttt{\textbackslash n}Line L12: condition is cond\_x (False), so skip L13-16.\texttt{\textbackslash n}\texttt{\textbackslash n}Line L17: executed, popping lst\_x's last element (8 → now 6 elements again? Wait, after L5, lst\_x was [9,3,5,2,6,8], then L17 pops the last (8), so becomes [9,3,5,2,6].\texttt{\textbackslash n}\texttt{\textbackslash n}Line L18: executed, appending v (2) to lst\_w.\texttt{\textbackslash n}\texttt{\textbackslash n}Line L19: executed, setting cond\_d to True.\texttt{\textbackslash n}\texttt{\textbackslash n}Line L20: condition is cond\_y (False), so skip L21.\texttt{\textbackslash n}\texttt{\textbackslash n}Line L22: return, so nothing else.\texttt{\textbackslash n}\texttt{\textbackslash n}So all steps accounted for. Now, let me check the lists again for each step.\texttt{\textbackslash n}\texttt{\textbackslash n}Wait, for L18's lst\_w:\texttt{\textbackslash n}\texttt{\textbackslash n}Original lst\_w is [2,8,2,1,7,9,9,5,8,5]. After appending v (2), the new list is:\texttt{\textbackslash n}\texttt{\textbackslash n}[2,8,2,1,7,9,9,5,8,5,2]. So that's correct.\texttt{\textbackslash n}\texttt{\textbackslash n}Now, the output should list each line where a variable was changed, with the line number and the variable and new value.\texttt{\textbackslash n}\texttt{\textbackslash n}Now, let me check the order of the lines. The lines are executed in order L2, L4, L5, L6, L7, L11, L12, L17, L18, L19, L20, L22.\texttt{\textbackslash n}\texttt{\textbackslash n}Wait, between L12 and L17, there's L12's if block, but since it's skipped, the next line is L17. So the order is correct.\texttt{\textbackslash n}\texttt{\textbackslash n}Now, the output should be:\texttt{\textbackslash n}\texttt{\textbackslash n}L2,\texttt{\textbackslash n}\texttt{\textbackslash n}L4,lst\_x:[9,3,5,2,6]\texttt{\textbackslash n}\texttt{\textbackslash n}L5,lst\_x:[9,3,5,2,6,8]\texttt{\textbackslash n}\texttt{\textbackslash n}L6,cond\_y:False\texttt{\textbackslash n}\texttt{\textbackslash n}L7,\texttt{\textbackslash n}\texttt{\textbackslash n}
    L11,lst\_z:[0,8,4,5,8,4,4,7]\texttt{\textbackslash n}\texttt{\textbackslash n}L12,\texttt{\textbackslash n}\texttt{\textbackslash n}L17,lst\_x:[9,3,5,2,6]\texttt{\textbackslash n}\texttt{\textbackslash n}L18,lst\_w:[2,8,2,1,7,9,9,5,8,5,2]\texttt{\textbackslash n}\texttt{\textbackslash n}
    L19,cond\_d:True\texttt{\textbackslash n}\texttt{\textbackslash n}L20,\texttt{\textbackslash n}\texttt{\textbackslash n}L22,\texttt{\textbackslash n}\texttt{\textbackslash n}Wait, but in the previous examples, sometimes there are more lines. Let me check if any steps were missed.\texttt{\textbackslash n}\texttt{\textbackslash n}Wait, after L11 comes L12, then L12's if is skipped, so L17 comes next. So yes, the order is correct.\texttt{\textbackslash n}\texttt{\textbackslash n}Wait, but in the input, the lst\_z after L11 is [0,8,4,5,8,4,4,7]. Let me confirm that:\texttt{\textbackslash n}\texttt{\textbackslash n}Original lst\_z was [0,8,4,5,8,4,4]. Appending 7 (w=7) gives the new list as [0,8,4,5,8,4,4,7]. That's correct.\texttt{\textbackslash n}\texttt{\textbackslash n}Another check: line L18 appends v=2 to lst\_w. The original lst\_w was [2,8,2,1,7,9,9,5,8,5], so adding 2 at the end gives the list as listed.\texttt{\textbackslash n}\texttt{\textbackslash n}Yes.\texttt{\textbackslash n}\texttt{\textbackslash n}Now, the final output should be as listed above. Let me format it exactly as per the examples, with commas separating entries and each line starting with L followed by the line number, then the variable and value if applicable.\texttt{\textbackslash n}\texttt{\textbackslash n}So the final answer should be:\texttt{\textbackslash n}\texttt{\textbackslash n}L2,\texttt{\textbackslash n}L4,lst\_x:[9,3,5,2,6]\texttt{\textbackslash n}L5,lst\_x:[9,3,5,2,6,8]\texttt{\textbackslash n}L6,cond\_y:False\texttt{\textbackslash n}L7,\texttt{\textbackslash n}
    L11,lst\_z:[0,8,4,5,8,4,4,7]\texttt{\textbackslash n}L12,\texttt{\textbackslash n}L17,lst\_x:[9,3,5,2,6]\texttt{\textbackslash n}L18,lst\_w:[2,8,2,1,7,9,9,5,8,5,2]\texttt{\textbackslash n}
    L19,cond\_d:True\texttt{\textbackslash n}L20,\texttt{\textbackslash n}L22,\texttt{\textbackslash n}\texttt{\textbackslash n}Wait, but in the previous examples, sometimes there are more entries. Let me check again if I missed any lines where variables are modified.\texttt{\textbackslash n}\texttt{\textbackslash n}Wait, line L19 is cond\_d = 1 != v. Since v is 2, 1 !=2 is True, so that's correct.\texttt{\textbackslash n}\texttt{\textbackslash n}Yes, that's all. I think that's the correct trace.\texttt{\textbackslash n}\texttt{<}/think\texttt{>}\texttt{\textbackslash n}\texttt{\textbackslash n}```\texttt{\textbackslash n}L2,\texttt{\textbackslash n}L4,lst\_x:[9,3,5,2,6]\texttt{\textbackslash n}L5,lst\_x:[9,3,5,2,6,8]\texttt{\textbackslash n}L6,cond\_y:False\texttt{\textbackslash n}L7,
    \texttt{\textbackslash n}L11,lst\_z:[0,8,4,5,8,4,4,7]\texttt{\textbackslash n}L12,\texttt{\textbackslash n}L17,lst\_x:[9,3,5,2,6]\texttt{\textbackslash n}L18,lst\_w:[2,8,2,1,7,9,9,5,8,5,2]\texttt{\textbackslash n}
    L19,cond\_d:True\texttt{\textbackslash n}L20,\texttt{\textbackslash n}L22,\texttt{\textbackslash n} \\
    \bottomrule
    \end{tabular}
    \label{tab:qwq_example_1}
\end{table}

\end{document}